\documentclass[pra,reprint,aps,amsmath,amssymb,twocolumn,floatfix,superscriptaddress]{revtex4-2}
\usepackage{amsmath}
\usepackage{graphicx}
\usepackage{color}
\usepackage{amssymb}
\usepackage[utf8]{inputenc}
\usepackage[T1]{fontenc}

\usepackage[normalem]{ulem}
\usepackage{array}
\usepackage{comment}
\usepackage{hyperref}
\hypersetup{colorlinks,%
linkcolor=blue,%
citecolor=blue,%
urlcolor=blue%
}

\begin{document}

\title{Supersolid phases and collective excitations in two-dimensional Rashba spin-orbit coupled spin-1 condensates}

\author{Sanu Kumar Gangwar}
\affiliation{Department of Physics, Indian Institute of Technology, Guwahati 781039, Assam, India} 

\author{Sayan Chatterjee}
\affiliation{Department of Physics, Indian Institute of Technology, Guwahati 781039, Assam, India} 

\author{Rajamanickam Ravisankar}
\affiliation{Department of Mechanics and Aerospace Engineering, Southern University of Science and Technology, Shenzhen 518055, China}

\author{Henrique Fabrelli}
\affiliation{Institute of Solid State Physics, TU Wien, Wiedner Hauptstr. 8-10, 1040 Vienna, Austria}

\author{Paulsamy Muruganandam}
\affiliation{Department of Physics, Bharathidasan University, Tiruchirappalli 620024, Tamilnadu, India}

\author{Pankaj Kumar Mishra}
\affiliation{Department of Physics, Indian Institute of Technology, Guwahati 781039, Assam, India}

\date{\today}
\begin{abstract} 
We investigate the collective excitation spectrum and dynamics of a quasi-two-dimensional spin-1 Bose-Einstein condensate with Rashba-type spin-orbit (SO) coupling. Employing Bogoliubov-de-Gennes analysis, we analytically compute the excitation spectra across a wide range of interaction strengths and coupling parameters. By systematically varying the SO and Rabi couplings, we uncover distinct dynamical signatures of quantum phase transitions, including mode softening, the appearance of roton-like minima, and miscibility-driven instabilities in both ferromagnetic and antiferromagnetic interaction regimes. In the antiferromagnetic case, these instabilities lead to a dynamically unstable supersolid phase characterized by the coexistence of density modulation and global phase coherence. To corroborate the analytical predictions, we numerically solve the coupled Gross–Pitaevskii equations and analyze the dynamical stability of the condensate. Our results provide experimentally accessible signatures for spinor condensates with tunable spin–orbit coupling and demonstrate the rich interplay between spin-dependent interactions and synthetic  couplings in nonequilibrium quantum fluids.
\end{abstract}

\maketitle
\section{Introduction}
\label{sec:1}

Ultracold gases provide a highly versatile and controllable platform for exploring many-body physics in the laboratory~\cite{bloch2008many, georgescu2014quantum, Safronova:2018, Chomaz:2022}, serving as a guiding framework for frontier quantum technologies~\cite{Krantz2019, Kaufman:2021}. The realization of Bose–Einstein condensates (BECs) enabled the study of macroscopic quantum coherence in dilute gases~\cite{Anderson1995, Davis1995, proukakis2025century}, which was later extended to spinor condensates in which atoms retain internal spin degrees of freedom~\cite{Stamper1998}. This internal spin structure has enabled the engineering of synthetic gauge fields, most notably spin–orbit (SO) coupling, which has been experimentally realized in quasi-one-dimensional condensates via Raman-induced equal Rashba–Dresselhaus coupling~\cite{Lin2011, Campbell2016, Luo2016} and in quasi-two-dimensional systems with Rashba-type coupling~\cite{Wu2016, Sun2018}. The incorporation of synthetic SO coupling into ultracold atomic gases has opened new avenues for engineering novel quasiparticle spectra and emergent quantum phases, including roton-like minima~\cite{Ji2015}, supersolid and stripe phases~\cite{Luo2019, Li2017Supersolid, Ho2011}, topological textures~\cite{Kawakami2012}, and Zitterbewegung oscillations~\cite{Qu2013}.

Numerical simulations of the mean-field Gross–Pitaevskii equations (GPEs) have played a central role in understanding SO coupled BECs, enabling detailed exploration of ground-state phases, spin textures, collective excitations, and nonequilibrium dynamics~\cite{pitaevskii2003bose}. In quasi-1D spin-1 BECs, SO coupling has been shown to induce rich physics, including stripe phases~\cite{chen2022elementary}, mode softening, and characteristic roton-like features in the excitation spectrum~\cite{Rajat2025}. In particular, the Bogoliubov analyses of Raman-induced SO-coupled spin-1 condensates revealed roton–maxon structures in the plane-wave and zero-momentum phases, where the softening of roton modes signals instabilities toward stripe and supersolid-like order~\cite{Yu2016, Sun2016}. Our previous works further demonstrated that in quasi-1D spin-1 condensates, SO coupling gives rise to dynamical instabilities manifested as unstable or double-unstable avoided crossings in ferromagnetic and antiferromagnetic phases, providing clear dynamical signatures of phase transitions~\cite{Gangwar2024, Gangwar2025, gangwar2025emergence}. Related studies also uncovered mode softening and double-roton structures across phase boundaries in quasi-1D systems at finite temperature~\cite{Rajat2025}.

In quasi-2D spin-1 systems, the interplay between SO coupling and interatomic interactions facilitates spontaneous density modulation and the emergence of supersolid-like phases~\cite{adhikari2021supersolid, Kaur2022}. Adhikari et al. demonstrated the formation of stable self-bound supersolid solitons in Rashba SO-coupled spin-1 condensates, characterized by stripe and superlattice density order~\cite{Adhikarimultr_2021, Adhikari2021127042}. This framework was later extended to nonmagnetic spin-1 and spin-2 condensates, where quasi-degenerate stripe and superlattice solitons remain stable over a wide range of SO coupling strengths~\cite{Kaur2022}. 

Collective excitations provide a powerful and effective probe of complex quantum phases, such as stripe and supersolid states. Their spectra encode essential information about dynamical stability, symmetry breaking, and quantum phase transitions, with the softening of roton modes serving as a hallmark of emerging density order~\cite{Bogoliubov1947, Jin1996, Mewes1996}. In SO coupled BECs, the interplay between Raman or Rashba coupling and interactions fundamentally reshapes the Bogoliubov spectrum, giving rise to anisotropic phonon modes, tunable roton minima, and interaction-driven instabilities~\cite{Higbie2002, Martone2012, Ravisankar2021, Ravisankar2025}.

Extensive theoretical studies have demonstrated that SO coupling can induce single or multiple roton structures, suppress long-wavelength phonon modes, and trigger dynamical instabilities in two-dimensional condensates and quantum droplets~\cite{Ozawa2012Stability, Ozawa2013, Sahu2020}. Moreover, SO coupling strongly modifies hydrodynamic responses and critical velocities, underscoring its profound impact on superfluid behavior~\cite{Zheng2012HydroSOC, Zhang2016superfluid}. Recent work on quasi-two-dimensional binary spin-$1/2$ condensates has further revealed that while Rabi coupling tends to stabilize the system, SO coupling generates intricate instability landscapes near phase boundaries, offering access to unconventional quantum phase transitions~\cite{Ravisankar2021,Ravisankar2025}.

Despite these advances, a comprehensive understanding of collective excitations in spin-1 SO-coupled condensates remains notably incomplete. In particular, homogeneous two-dimensional systems with Rashba SO coupling, where, the competition between spin-dependent interactions, higher-spin degrees of freedom, and synthetic gauge fields is most pronounced, have received comparatively little attention~\cite{Sun2016, Sun2018}. As a result, the excitation spectra and dynamical stability of superstripe and supersolid phases in spin-1 condensates, especially across both ferromagnetic and antiferromagnetic interaction regimes, are not well explored. Clarifying how these complex phases manifest in the collective modes is essential for identifying experimentally accessible signatures and for distinguishing stable supersolids from dynamically unstable stripe states~\cite{adhikari2021supersolid, Adhikari2021127042}.

In this paper, we address this gap by performing a systematic Bogoliubov analysis of a uniform quasi-two-dimensional spin-1 Bose–Einstein condensate with Rashba SO and Rabi couplings. We compute the collective excitation spectra across a wide range of interaction parameters, encompassing both ferromagnetic and antiferromagnetic regimes. Our analysis reveals distinct mode softening, roton minima, and miscibility-driven instabilities that serve as clear dynamical signatures of quantum phase transitions. Notably, we demonstrate that supersolid phases in this system are generically dynamically unstable, while superstripe phases exhibit characteristic excitation features. Our results may be useful to observe these supersolid and superstripe behavior through collective excitation measurements experimentally in SO coupled spinor condensates.

The paper is structured as follows. In Sec.~\ref{sec:model}, we introduce the mean-field model and describe the numerical framework used to investigate collective excitations. The single-particle spectrum is analyzed in Sec.~\ref{sec:singpart}, laying the groundwork for the derivation of the collective excitation spectrum in Sec.~\ref{sec:collexc}. In Sec.~\ref{sec:stabphase}, we present the stability phase diagram for ferromagnetic interactions, followed by the stability analysis with Rabi coupling (Sec.~\ref{sec:5a}) and without Rabi coupling (Sec.~\ref{sec:5b}). The stability analysis for antiferromagnetic interactions is given in Sec.~\ref{sec:stabphaseaferro}. Finally, in Sec.~\ref{sec:summ}, we offer the summary and conclusions of our findings. A detailed derivation of the Bogoliubov–de Gennes (BdG) matrix is provided in Appendix~\ref{matrx:BdG}. In the Appendix~\ref{afm:secp}, we present the different phases that arise for different combinations of the SO and Rabi couplings for anti-ferromagnetic interactions.

\section{Mean-Field Model}
\label{sec:model}
We consider a harmonically trapped quasi-two-dimensional (quasi-2D) spin-1 condensate with SO coupling realized via the tight confinement along the z-direction ($\omega_z \gg \omega$), where $\omega_z$ and $\omega$ are the angular trap frequencies in the $z$-direction and in the $x-y$ plane, respectively. Within this framework, the dimensionless coupled Gross–Pitaevskii equations (GPEs) governing the dynamics of the SO-coupled spin-1 condensate are given by~\cite{Kawaguchi2012, Yukalov_2018, Ravisankarcpc2021, Adhikari_2021},
\begin{subequations}
\begin{align}
\mathrm{i} \frac{\partial}{\partial t} \psi_{\pm 1}(\vec{r}) & = \left[-\frac{1}{2} \nabla^{2} + V(\vec{r}) + c_{0} \rho + c_{2} \rho_2^{\pm} \right] \psi_{\pm 1}(\vec{r}) \notag \\  & + c_{2} \psi_{0}^{2} (\vec{r}) \psi_{\mp 1}^{*} (\vec{r}) + \frac{\Omega}{\sqrt{2}} \psi_{0}(\vec{r}) \notag \\  & - \mathrm{i} \frac{k_{L}}{\sqrt{2}} \left[ \bigg(\frac{\partial}{\partial y} \pm \mathrm{i} \frac{\partial}{\partial x}\bigg)  \psi_{0}(\vec{r})\right], \label{gpe1} \\
\mathrm{i} \frac{\partial}{\partial t} \psi_{0}(\vec{r}) &= \left[-\frac{1}{2} \nabla^{2} + V(\vec{r}) + c_{0} \rho + c_{2} \rho_{2}^{0} \right] \psi_{0}(\vec{r}) \notag \\ & + 2 c_{2} \psi_{+1}(\vec{r}) \psi_{-1}(\vec{r}) \psi_{0}^{*}(\vec{r}) + \frac{\Omega}{\sqrt{2}}(\psi_{+}(\vec{r})) \notag \\ & - \mathrm{i} \frac{k_{L}}{\sqrt{2}} \left[
\frac{\partial}{\partial y} \psi_{+}(\vec{r}) - \mathrm{i} \frac{\partial}{\partial x} \psi_{-}(\vec{r})\right], \label{gpe2}
\end{align} 
\end{subequations}
where $\vec{r} =\{ x, y\}$, $\psi_{j}$; $j=+1,0,-1$, are the spinor condensate wavefunctions corresponding to sublevels $m_{F} = +1, 0, -1$ of the hyperfine state $F = 1$ that satisfy the normalization condition $\int_{-\infty}^{\infty} \rho \; d x dy= 1$, where $\rho_j= \vert\psi_{j}\vert^{2}$, and $\rho=\vert\psi_{+1}\vert^{2} + \vert\psi_{0}\vert^{2} + \vert\psi_{-1}\vert^{2}$. Also, $\psi_{\pm}=\psi_{+1} \pm \psi_{-1}$, $\rho_2^{\pm}=\left(\rho_{\pm1} + \rho_{0} - \rho_{\mp1}\right) $, $\rho_2^0=\left( \rho_{+1} + \rho_{-1}\right)$. The equations (\ref{gpe1}), and (\ref{gpe2}) are non-dimensionalized using time, lengths, and energy $t = \omega \tilde{t}$, $x = \tilde{x}/l_{0}$, $y = \tilde{y}/l_{0}$, and $\hbar \omega$, respectively. The resulting condensate wavefunction is $\psi_{\pm 1,0} = \frac{l_{0}}{\sqrt{N}} \tilde{\psi}_{\pm 1,0}$, where, $l_{0} = \sqrt{\hbar/ m \omega}$, is the oscillator length for the trap frequency $\omega$ in the $x$-$y$ plane. The trap strength is given by $V (x,y) = (x^{2} + y^{2}) /2$, density-density interaction strength $c_{0} = 2 N \sqrt{2 \pi \kappa}(a_{0} + 2a_{2})/ 3 l_{0}$, and spin-exchange interaction strength $c_{2} = 2 N \sqrt{2 \pi \kappa}(a_{2} - a_{0})/ 3 l_{0}$, where $\kappa = \omega_{z}/ \omega$, $a_{0}$ and $a_{2}$ are the s-wave scattering lengths in total spin channels $0$ and $2$, respectively. Upon tuning $c_{2} < 0$ ($ c_2> 0$), one can have the FM (AFM) interaction of the condensates~\cite{Kawaguchi2012}. The SO and Rabi coupling strengths are given by  $k_{L} = \tilde{k}_{L} / \omega l_{0}$, $\Omega = \tilde{\Omega} / (\hbar \omega)$, respectively. In the above description, the quantities with the tilde represent dimensionful quantities. In this entire work, we consider all parameters to be dimensionless.%

The energy functional corresponding to the coupled GPEs (\ref{gpe1})-(\ref{gpe2}) is given by~\cite{Ravisankarcpc2021, Adhikari_2021},
\begin{align}
E  = & \frac{1}{2} \int dxdy \bigg\{ \sum_{j} (\left\vert \partial_{x}\psi_{j} \right\vert^{2} +  \left\vert \partial_{y}\psi_{j} \right\vert^{2})  + 2 V(x,y) \rho   \nonumber  \\  & + c_{0} \rho^{2} + c_{2}[ \rho_{+1}^{2} + \rho_{-1}^{2} + 2( \rho_{+1}\rho_{0}+\rho_{-1}\rho_{0}  \nonumber  \\  & -\rho_{+1}\rho_{-1}+\psi_{-1}^{*}\psi_{0}^{2}\psi_{+1}^{*}+ \psi_{-1}\psi_{0}^{*2}\psi_{+1})]   \nonumber  \\  & + \sqrt{2} \Omega[\psi_{+}^{*}\psi_{0}+\psi_{0}^{*}\psi_{+}] -\sqrt{2} \mathrm{i}k_{L}[\psi_{+}^{*} \partial_{y}\psi_{0}  \nonumber \\ & +\mathrm{i}\psi_{-}^{*}\partial_{x}\psi_{0} - \mathrm{i} \psi_{0}^{*} \partial_{x}\psi_{-} + \psi_{0}^{*} \partial_{y} \psi_{+} ]\bigg\} \label{eqn5}
\end{align}%

\begin{figure*}[!htb]
\centering\includegraphics[width=0.99\linewidth]{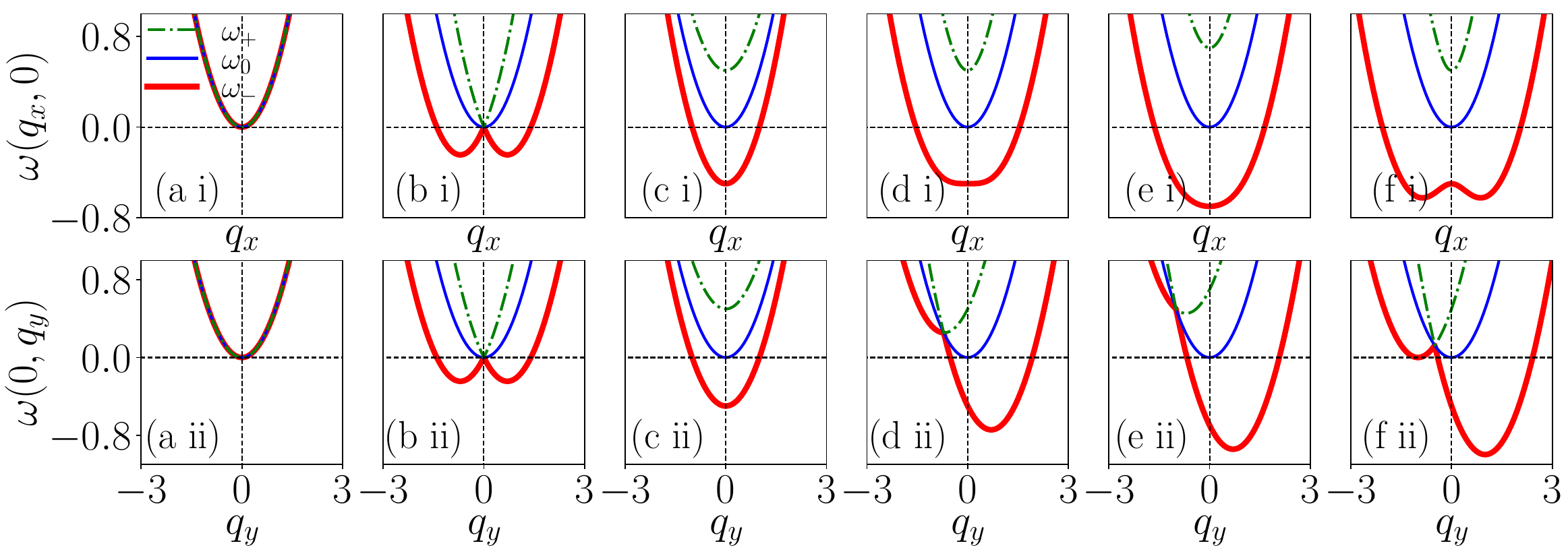}
\caption{Single-particle energy spectrum top row along $q_{x}$ direction and bottom row along $q_{y}$ direction for the different set of SO and Rabi coupling strengths $(k_L,\Omega)$: (a)-(f): $(0, 0)$, $(0.7, 0)$, $(0, 0.5)$, $(0.7, 0.5)$, $(0.7, 0.7)$, $(1.0, 0.5)$. The thick solid red, thin solid blue line, and dash-dotted green line represent the energy eigenspectrum for $\{-1, 0, +1\}$ components of the spin, respectively. Along $q_{x}$ direction, the double minimum appears in the eigenvalue dispersion, following the relation $k_{L}^{2} > \Omega$, which shows zero-momentum to stripe-wave phase transitions. While along the $q_{y}$ direction, the eigenvalue dispersion exhibits the asymmetric double minima in the presence of both SO and Rabi coupling.}
\label{fig1} 
\end{figure*}
Here, we outline the experimentally feasible parameter ranges relevant to our simulations. For the ferromagnetic interaction, we consider $^{87}$Rb BECs with $N \sim 3\times 10^{5}$ atoms confined in a harmonic trap with the frequencies ($\omega, \omega_{z}$) = $2 \pi \times$ ($20 \text{Hz}, 600 \text{Hz}$). The corresponding characteristic lengths would be $l_{0} = 2.4 \mu$m, and $l_{z} = 0.44 \mu$m. For the antiferromagnetic interaction, we consider the BECs of $^{23}$Na atoms, for which the characteristic lengths are $l_{0} = 4.7 \mu$m, and $l_{z} = 0.85 \mu$m~\cite{Adhikari_2021, Kazuya2012}.  The spin-dependent and spin-independent interactions can be achieved by controlling the $s$-wave scattering lengths through Feshbach resonance~\cite{Inouye1998, Marte2002, Chin2010}. The SO coupling strength $k_L = \{ 0.1 - 5 \}$ can be tuned by changing the laser wavelengths in the range of $\{151.19 \mu$m - $3023.89 \mbox{nm}\}$. However, the dimensionless Rabi frequency interval $\Omega=[0, 11]$ used in the simulation can be attained by tuning the Raman laser strength in the range of  $2\pi \hbar\times \{5 - 220\}$ Hz.%

\section{Single-particle spectrum}
\label{sec:singpart}
In this section, we present the single-particle spectrum of the non-interacting spinor condensate for trapless SO and Rabi-coupled spin-1 spinor BECs. Following that, we introduce the collective excitation spectrum for the interacting systems.

For $V(x, y) = 0$, and $ c_{0} =  c_{2} = 0$ and substituting plane wave solution, $\psi_{0,\pm 1} = \phi_{0,\pm 1} e^{\mathrm{i}({q_{x} x + q_{y} y -\omega t})}$ in the Eqs.(\ref{gpe1})-(\ref{gpe2}) we obtain,
\begin{align}\label{eqn1}
\mathcal{L}_{sp}\begin{pmatrix}
           \phi_{1}\\
     \phi_{0}\\
     \phi_{-1}
\end{pmatrix} & = \omega\begin{pmatrix}
     \phi_{1}\\
     \phi_{0}\\
     \phi_{-1}
     \end{pmatrix} 
\end{align}
with
\begin{align}
\mathcal{L}_{sp} =
 \begin{pmatrix}
 \frac{Q^{2}}{2} & \mathcal{L}_{12} & 0\\
 \mathcal{L}_{21}& \frac{Q^{2}}{2}  & \mathcal{L}_{23} \\
 0 & \mathcal{L}_{32}  & \frac{Q^{2}}{2}  \\
 \end{pmatrix},
\end{align} 
where, $Q^{2} = q_{x}^{2} + q_{y}^{2}$, $\mathcal{L}_{12} = \frac{\Omega}{\sqrt{2}} +\frac{k_{L}}{\sqrt{2}}(q_{y} + \mathrm{i} q_{x})$, $\mathcal{L}_{21} = \frac{\Omega}{\sqrt{2}} -\frac{k_{L}}{\sqrt{2}}(\mathrm{i}q_{x} - q_{y})$, $\mathcal{L}_{23} = \frac{\Omega}{\sqrt{2}} + \frac{k_{L}}{\sqrt{2}}(\mathrm{i}q_{x} + q_{y})$, and $\mathcal{L}_{32} = \frac{\Omega}{\sqrt{2}}- \frac{k_{L}}{\sqrt{2}}(-q_{y} + iq_{x})$.  Upon diagonalizing the Eq.(\ref{eqn1}), we obtain the single-particle dispersion as,
\begin{subequations}\label{eqn2}
\begin{align}%
\omega_{0}(q_{x}, q_{y})= &\frac{Q^{2}}{2} \\
\omega_{\pm}(q_{x}, q_{y})=&\frac{1}{2}\bigg( Q^{2} \pm 2 \sqrt{ k_{L}^{2} Q^{2} + 2 q_{y} k_{L} \Omega +  \Omega^{2}}\bigg)
\end{align}
\end{subequations}%


The structure of the single-particle spectrum in general provides direct insight into how SO and Rabi couplings modify the underlying energy landscape and thereby control the ground-state momentum of the system. In particular, the appearance or disappearance of degenerate minima in the lower branch determines whether the system favors a zero-momentum or finite-momentum state which directly dictate different ground state phases of the condensate~\cite{Lyu2020, Ravisankar2021, gangwar2025emergence}. In the single particle spectrum, we obtain three distinct branches, namely, $\omega_{+}$, $\omega_{0}$, and $\omega_{-}$. Among these, the $\omega_{0}$ branch remains independent of the SO and Rabi couplings, whereas the $\omega_{+}$ and $\omega_{-}$ exhibit explicit dependence on the SO and Rabi couplings. To simplify the analysis, we first consider the case $q_{y} = 0.0$ and examine the spectrum as a function of $q_{x}$. we then reverse the scenario by fixing $q_{x} = 0.0$ and exploring the variation with respect to $q_{y}$.

\paragraph*{In $q_{x}$ direction.} The top row of Fig.\ref{fig1} shows the single-particle dispersion along the $q_{x}$ direction for $q_{y} = 0$. In the absence of the SO and Rabi couplings ($k_{L} = \Omega = 0$)  the spectrum consists of a non-degenerate parabolic dispersion band [see Fig.\ref{fig1}(a i)], in line with the expression in Eq.~\ref{eqn2}. Introducing SO coupling alone ($k_{L} = 0.7$) qualitatively alters the dispersion: the lower branch $\omega_{-}$ develops two degenerate minima at finite momenta $q_{x}=\pm 0.7$ as shown in the Fig.~\ref{fig1}(b i), signaling the tendency toward finite-momentum states induced by SO coupling.  In contrast, when only the Rabi coupling is present ($k_{L}=0$, $\Omega=0.5$), the degeneracy is lifted and a gap opens between the branches [Fig.~\ref{fig1}(c i)]. The energy separation between $\omega_{0}$ and $\omega_{\pm}$ is $\Omega$, while the gap between $\omega_{+}$ and $\omega_{-}$ equals $2\Omega$, reflecting coherent spin mixing due to the Rabi term. As a result, the lower branch exhibits a single minimum at zero momentum.

When both SO couplings are present ($k_{L} = 0.7$, $\Omega = 0.5$), the competition between momentum-dependent SO effects and spin mixing leads to a partial flattening of the minimum of the $\omega_{-}$ branch [see Fig.~\ref{fig1}(d i)]. Increasing the Rabi coupling further ($\Omega=0.7$) suppresses this flattening and restores a single minimum [Fig.~\ref{fig1}(e i)]. However, for stronger SO coupling ($k_{L}=1.0$) at fixed $\Omega=0.5$, the lower branch again develops a clear double minimum despite the presence of a finite gap as illustrated in Fig.~\ref{fig1}(f i). The emergence of a double minimum corresponds to a transition from the zero-momentum (ZM) phase to the stripe-wave (SW) phase, governed by the condition $k_{L}^{2}>\Omega$.


\paragraph*{In $q_{y}$ direction.}

We now turn to the dispersion along the $q_y$ direction by fixing $q_x=0$, as shown in the bottom row of Fig.~\ref{fig1}. Here we find the modification in the momentum dependence of the spectrum which leads to qualitatively distinct behavior compared to the $q_{x}$ direction. In the absence of both couplings, or when only one of them is present, the dispersions in panels Fig.~\ref{fig1}(a ii–c ii) closely resemble their counterparts along the $q_{x}$ direction [Fig.~\ref{fig1}(a i–c i)]. In particular, finite SO coupling in the absence of Rabi coupling produces a symmetric double minimum in the lower branch $\omega_{-}$, reflecting degenerate finite-momentum states.

A qualitatively new feature emerges when both the SO and Rabi couplings are finite. For $k_{L}=0.7$ and $\Omega=0.5$ [Fig.~\ref{fig1}(d ii)], the lower branch develops an asymmetric double-minimum structure. The global minimum occurs at $q_{y}=0.7$ with energy $\omega_{-}=-0.745$, while the secondary minimum lies at positive energy. Increasing the Rabi coupling to $\Omega=0.7$ further deepens the global minimum to $\omega_{-}=-0.945$ [Fig.~\ref{fig1}(e ii)], enhancing the asymmetry between the two minima. In both cases, the two minima are energetically inequivalent, indicating a lifting of the degeneracy between opposite momenta.

For stronger SO coupling, $k_{L}=1.0$ with $\Omega=0.5$ [Fig.~\ref{fig1}(f ii)], the asymmetric double-minimum structure persists. One minimum appears at zero energy ($\omega_{-}=0$ at $q_{y}=-1.0$), while the other remains at negative energy ($\omega_{-}=-1.0$ at $q_{y}=1.0$), demonstrating a pronounced imbalance between positive and negative momenta. Overall we find that while symmetric double minima arise solely from SO coupling and correspond to degenerate finite-momentum states, the simultaneous presence of SO and Rabi couplings breaks this symmetry and selects a preferred momentum direction. This asymmetry reflects the anisotropic nature of the SO coupling and has important consequences for momentum selection and ground-state structure~\cite{Ravisankar2021, Sun2016}.

\section{Collective Excitation spectrum}
\label{sec:collexc}
In the previous section, we analyzed the single-particle spectrum of SO coupled spin-1 BECs, which determines the underlying band structure and ground-state momentum configurations. However, the stability and dynamical response of the condensate are governed by collective excitations, which crucially depend on interatomic interactions. Within this approach, the condensate wave function is expressed as a small fluctuation about the uniform mean-field ground state. The excitation wave function for the $j$th spin component can be written as~\cite{Ravisankar2021,Zhu2020},
\begin{align}\label{excwave}
\psi_{j}(x,y,t) = \mathrm{e}^{-\mathrm{i} \mu_{j} t}[\phi_{j} + \delta \phi_{j}(x, t) ]
\end{align}    
where, where $\phi_{j}$ denotes the uniform ground-state wave function, $\mu_{j}$ is the chemical potential, and $\delta \phi_{j}$ represents a small time-dependent perturbation.

The fluctuation term is expanded in plane-wave modes as
\begin{align}\label{pertwave}
\delta \phi_{j}(x,y,t) = u_{j} \mathrm{e}^{\mathrm{i}( q_{x} x + q_{y} y - \omega t)} + v_{j}^{*} \mathrm{e}^{- \mathrm{i}(q_{x} x + q_{y} y - \omega^{*} t)}
\end{align}%
where, $u_{j}$,  $v_{j}$ are the Bogoliubov amplitudes and $(q_{x},q_{y})$ denotes the excitation momentum. The ground state of the spinor is taken as $\phi_{j} = (1/2, -1/\sqrt{2}, 1/2)^{T}$ corresponding to the uniform density. Here, $\mu_{j}$ denotes the chemical potential. The index $j=+1,0,-1$ labels the three spin components of the spin-1 condensate. Substituting Eq.~(\ref{excwave}) into the coupled Gross–Pitaevskii equations [Eqs.~(\ref{gpe1})–(\ref{gpe2})] and retaining terms up to linear order in the fluctuations, we obtain the BdG equations governing the collective excitation spectrum as given by
\begin{widetext}
 \begin{align}\label{bdgeigprbm}
    \mathcal{L} \begin{pmatrix}
            u_{+1} & v_{+1} & u_{0} & v_{0} & u_{-1} & v_{-1}\\
          \end{pmatrix}^{T}
     &=\omega \begin{pmatrix}
     u_{+1} & v_{+1} & u_{0} & v_{0} & u_{-1} & v_{-1}\\
     \end{pmatrix}^{T} 
   \end{align}
where $T$ represents the transpose of the matrix and  $\mathcal{L}$ is $6 \times 6$ matrix given by,
\begin{align} \label{bdgmatrix}
 \mathcal{L} = 
 \begin{pmatrix}
 H_{+}-\mu_{+} & \mathcal{L}_{12} & \mathcal{L}_{13}
 & \mathcal{L}_{14} & \mathcal{L}_{15} &
 \mathcal{L}_{16}\\
 \mathcal{L}_{21} & -H_{+}+\mu_{+} & \mathcal{L}_{23} & \mathcal{L}_{24}
& \mathcal{L}_{25} & \mathcal{L}_{26}\\
 \mathcal{L}_{31} & \mathcal{L}_{32} & H_{0}-\mu_{0} & \mathcal{L}_{34} & \mathcal{L}_{35} & \mathcal{L}_{36}\\
 \mathcal{L}_{41} & \mathcal{L}_{42} & \mathcal{L}_{43} & -H_{0}+\mu_{0} & \mathcal{L}_{45} & \mathcal{L}_{46}\\
 \mathcal{L}_{51} & \mathcal{L}_{52} & \mathcal{L}_{53} & \mathcal{L}_{54} & H_{-} -\mu_{-} & \mathcal{L}_{56}\\
 \mathcal{L}_{61} & \mathcal{L}_{62} & \mathcal{L}_{63} & \mathcal{L}_{64} & \mathcal{L}_{65} & -H_{-}+\mu_{-}
 \end{pmatrix}
 \end{align}
 \end{widetext}
The matrix elements of $\mathcal{L}$ are given in appendix~\ref{matrx:BdG}. Bogoliubov coefficients follow the normalization condition,
\begin{align}\label{normbdg}
\int \bigg(\sum_{j}\{\lvert u_{j} \rvert^{2} - \lvert v_{j} \rvert^{2}\} \bigg) dx dy = 1
\end{align}
The simplified form of the BdG excitation spectrum is obtained by calculating the determinant of the matrix $\mathcal{L}$ and equating it with zero, i.e., $det~(\mathcal{L} - \omega \mathrm {I}) = 0$, where $\mathrm{I}$ is a $6 \times 6$ identity matrix. The characteristic equation can be written as,
\begin{align}\label{bdgex}
 \omega^{6} + b \omega^{5} + c \omega^{4} + d \omega^{3} + e \omega^{2} + f \omega + g =0
\end{align}
where the coefficients $b$, $c$, $d$, $e$, $f$, and $g$ are given in the appendix~\ref{matrx:BdG}.%

\section{Stability phases based on collective excitation for Ferromagnetic interactions}
\label{sec:stabphase}

\begin{figure}[!htp]
\centering\includegraphics[width=0.90\linewidth]{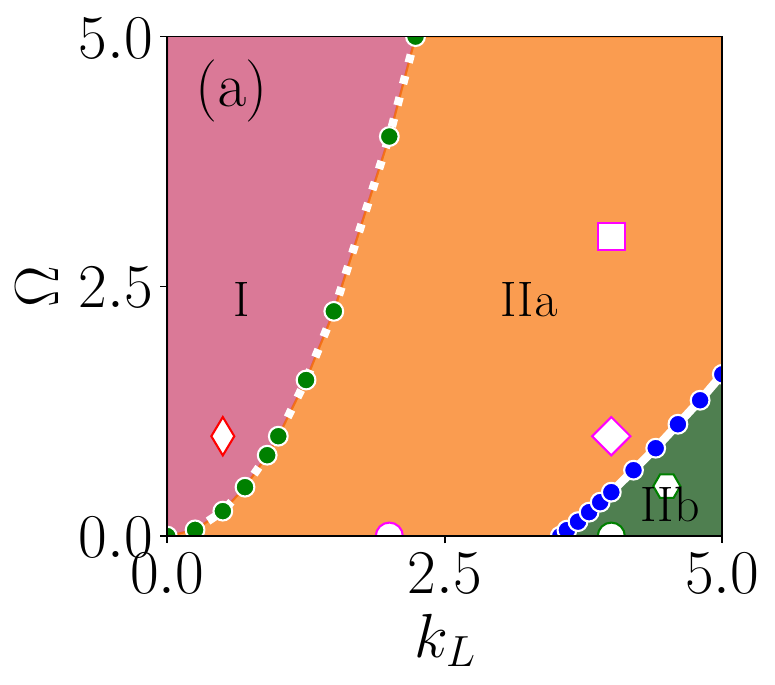}
\caption{Stability phase diagram in the $k_{L}$–$\Omega$ plane for ferromagnetic interactions ($c_{0} = 50.0$ and $c_{2} = -2.5$). (a) Full range of $q_x$ with $q_y=0$. The dash--dotted white line with green dots ($k_L^2\simeq\Omega$) separates stable (I) and unstable (II) regions. Within region II, the solid white line with blue dots ($\Omega=-1.6329+0.13k_L^2$) distinguishes gapped (IIa) and gapless (IIb) spectra. The case $\Omega \sim 0$ lies in the unstable regime. The marker in different regions represents the coupling parameters points for which detailed collective excitation analysis has been reported in the paper.}
\label{fig3} 
\end{figure}

The dynamical and energetic stability of the condensate is determined from the associated eigenfrequencies. The condensate is both dynamically and energetically stable when all eigenfrequencies are real and positive. The appearance of complex eigenfrequencies signals dynamical instability, while purely real but negative eigenfrequencies indicate energetic instability~\cite{Goldstein1997, Ozawa2013, Ravisankar2021}. Fig.~\ref{fig3} presents the stability phase diagram in the $k_{L}-\Omega$ plane in presence of ferromagnetic interactions with interaction strengths $c_{0} = 50.0$ and $c_{2}= -2.5$ along $q_{x}$ at $q_{y} = 0.0$.

Based on the eigenspectrum in Fig.~\ref{fig3}, we divide the $k_{L} - \Omega$ phase plane  into three different regions: region \text{I}, region \text{II} (further divided into the regions \text{IIa} and \text{IIb}), and region \text{III}. The white dashed-dotted line with green dots follows the criterion $k_{L}^{2} \approx \Omega$ and separates region I and region II. Region I exhibits only real, positive eigenfrequencies and is both dynamically and energetically stable; region II contains a complex eigenspectrum and is dynamically unstable. Further, the solid white line with blue dots separates region II into two parts: IIa and IIb, which holds the relation $\Omega = -1.6329 + 0.13 k_{L}^{2}$ with origin point ($k_{L} = 3.54$, $\Omega \sim 0$). We obtain the gapped mode between the branches in IIa, and the gapless mode between the branches in region IIb. The case with vanishing Rabi coupling ($\Omega \sim 0.0$) is treated separately as region III. This region is dynamically unstable, supporting both single-band instabilities with a gapped mode and multi-band instabilities with gapless modes. For larger SO coupling ($k_{L} \gtrsim 4.1$), region III also exhibits phonon modes together with instability bands in the low-lying and first-excited branches which will discussed in more detail at the later part of the manuscript.

\begin{figure*}[!htp]
\centering\includegraphics[width=0.99\linewidth]{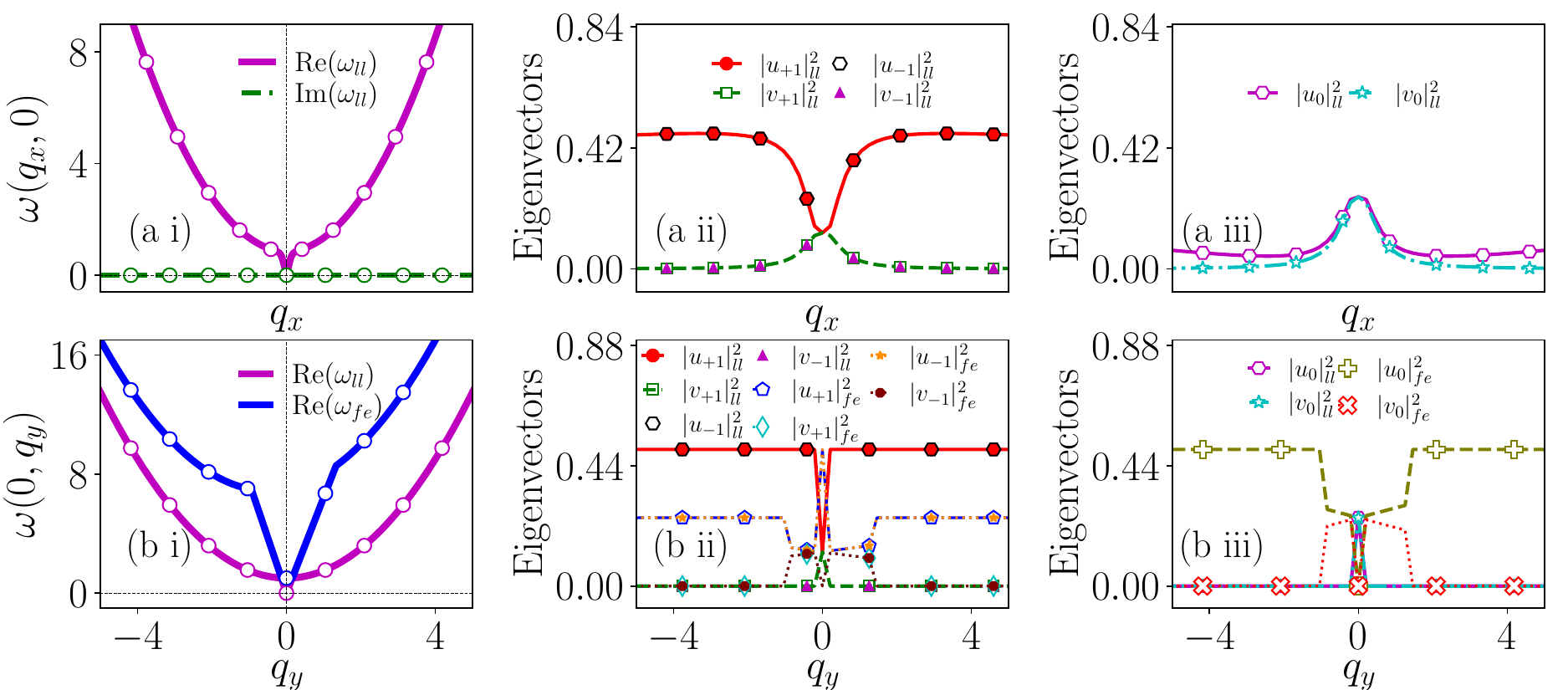}
\caption{Eigenvalue spectrum and eigenvectors along the quasi-momentum directions $q_{x}$ and $q_{y}$ for $k_{L} = 0.5$, $\Omega = 1.0$, $c_{0} = 50$, and $c_{2} = -2.5$.  Top row: low-lying brach. (a i) Eigenvalue spectrum, with $\mathrm{Re}(\omega_{ll})$ (solid magenta) and $\vert{\mathrm{Im}(\omega_{ll})}\vert$ (dashed dotted green). (a ii, a iii) Corresponding eigenvector components $\vert u_{+m}\vert^{2}_{ll}$ and $\vert v_{m}\vert^{2}_{ll}$ for $m=0,\pm1$. Bottom row: first excited branch. (b i) Eigenvalue spectrum with $\mathrm{Re}(\omega_{fe})$ (solid blue). (b ii, b iii) Corresponding eigenvector components  $\vert u_{m}\vert^{2}_{fe}$ and  $\vert u_{m}\vert^{2}_{fe}$. All eigenvalues are real and positive, and the low-lying branch exhibits a phonon mode, indicating both dynamical and energetic stability.  }
\label{fig08a} 
\end{figure*}
To analyze these features in more quantitative way, we also calculate the excitation spectrum analytically and employ numerical solution of the BdG equation [Eq.~\ref{bdgmatrix}]. In addition we also compute the corresponding eigenvectors for the eigenvalue spectrum along the quasi-momentum $q_{x}$ direction at $q_{y} = 0$, and along the $q_{y}$ direction at $q_{x} = 0$. First, we consider a real space grid $[-300:300][-300:300]$ with step size $h_{x} = h_{y} = 0.05$. We then apply the Fourier collocation method, where we numerically compute the Fourier transform of the BdG equations and obtain a truncated reduced BdG matrix, which is subsequently diagonalized using the LAPACK package~\cite{Anderson1999}. In momentum space, we consider $[-50:50] [-50:50]$ modes in the $q_x$, $q_{y}$ directions, with a grid step size of $h_{q_x}= h_{q_y} = 0.21$.%
Following this, we present the time evolution of the condensate associated with the collective excitation spectrum of the coupled Gross-Pitaevskii equations (GPEs) (Eqs.(\ref{gpe1})-(\ref{gpe2})). The ground state is obtained using the imaginary-time propagation method and subsequently evolved using the real-time propagation method. For this purpose, we employ the split-step Crank-Nicholson scheme described in Refs.~\cite{Muruganandam2009, Luis2016, Ravisankarcpc2021}. We have considered a grid of $N_{x} = N_{y} = 160$ space points with space step $dx = dy = 0.1$, time step $dt = 0.001$ for imaginary-time-propagation, and $dt = 0.00025$ for real-time-propagation. Initially, we obtain the ground states according to the regions in the phase plot [Fig.~\ref{fig3}] by using imaginary-time propagation. After calculating the ground state, we evolve it using real-time propagation by quenching the trap strength.%

Since the $k_{L} -\Omega$ phase plane is divided into three distinct regions: regions I, II (subdivided into IIa and IIb), and III, in the following subsections, we analyze the condensate behaviour by separating it into two cases: (A)in the presence of Rabi coupling (regions I and II), and (B) in the absence of Rabi coupling (region III). For each case, we report the collective excitations and numerical simulations using mean-field GP equation.%

\subsection{Effect of finite SO and Rabi coupling on collective excitation spectrum}
\label{sec:5a}
In the previous section, we presented the stability phase diagram, which is divided into two distinct regimes for finite Rabi coupling : I, II (subdivided into IIa and IIb), while, in the absence of Rabi coupling strength, there is a single region, III. In what follows we present the detailed analysis of eigen-spectrum and eigenvectors in the these regions.

\subsubsection{Region I} 
\label{sec:5ai}
We begin by considering the coupling parameters $(k_{L}, \Omega) = (0.5, 1.0)$, which lie within region I of the stability phase diagram shown with diamond marker in Fig.~\ref{fig3}. For this choice of coupling strengths, we analyze both the collective excitation spectrum and the corresponding numerical simulations.

\paragraph{Collective excitation spectrum:}
Figure~\ref{fig08a} displays the collective excitation spectra for the regime when $k_L^2<\Omega$, i.e.,  for $(k_{L}, \Omega) = (0.5, 1.0)$ with interaction strengths $c_{0} = 50$ and $c_{2} = -2.5$, plotted along the quasi-momentum directions $q_{x}$ (top row) and $q_{y}$ (bottom row). In the top row, Fig.~\ref{fig08a}(a i) shows that the low-lying branch of the spectrum consists solely of real and positive eigenfrequencies and supports a phonon mode. For a selected branch of the eigenspectrum, we examine the eigenvector components $\vert u_{\pm 1} \vert^{2}$, $\vert v_{\pm 1} \vert^{2}$, $\vert u_{0} \vert^{2}$, and $\vert v_{0} \vert^{2}$. As shown in Fig.~\ref{fig08a}(a ii), the eigenvectors exhibit in-phase behavior, characteristic of a density-like mode, satisfying the condition~\cite{Ravisankar2021, Gangwar2025},
\begin{align} \label{eqa:density}
& \vert u_{+1} \vert_{ll}^{2} - \vert u_{-1} \vert_{ll}^{2} = 0, \quad
\vert v_{+1} \vert_{ll}^{2} - \vert v_{-1} \vert_{ll}^{2} = 0.
\end{align}
The zeroth component, shown in Fig.~\ref{fig08a}(a iii), also exhibits independent in-phase behavior. Moreover, in the long-wavelength limit $q_{x} \approx 0$, all eigenvector components converge to the same value, confirming the phonon character of the low-lying excitation.
\begin{figure}[!htp]
\centering\includegraphics[width=0.99\linewidth]{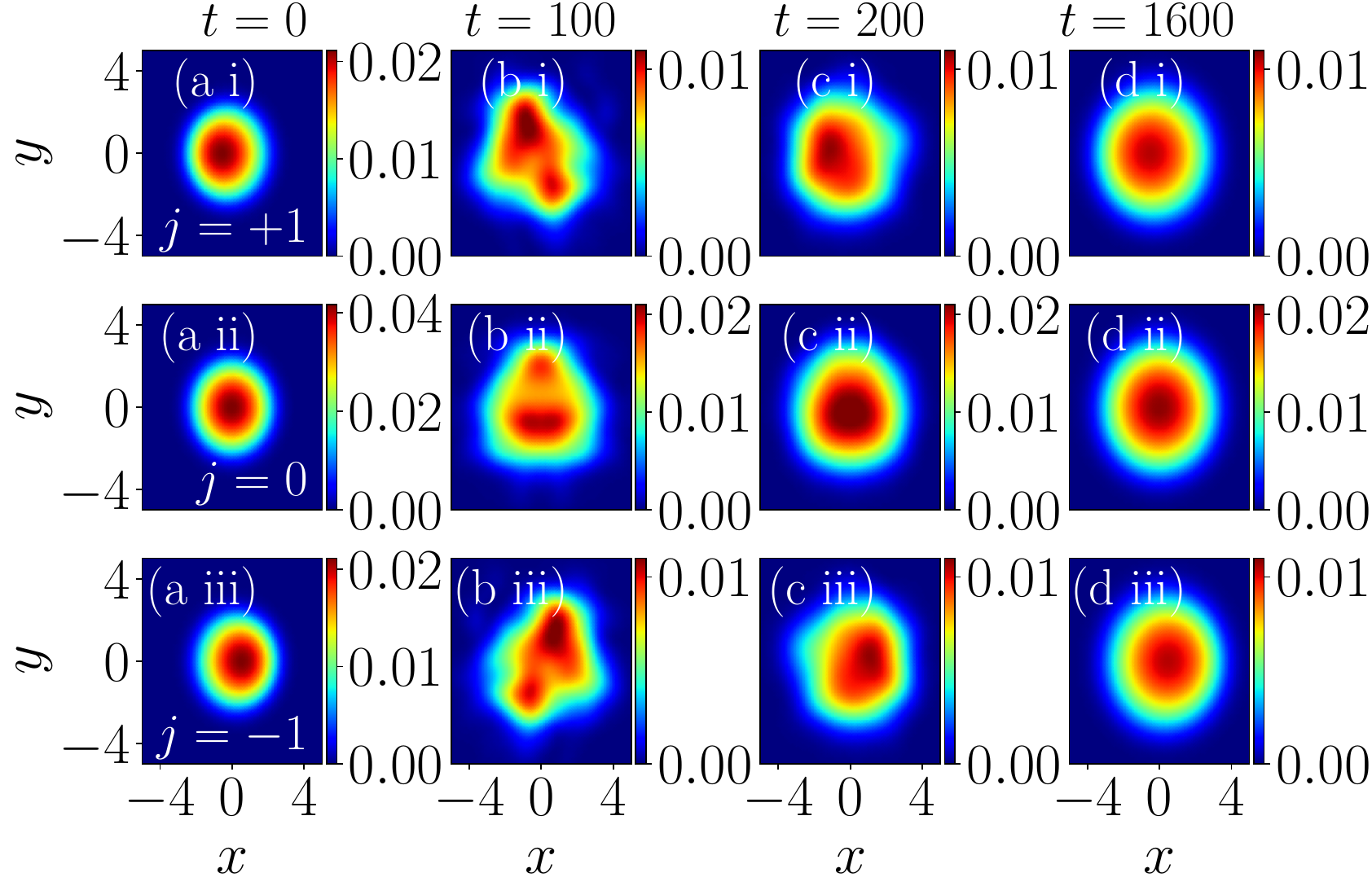}
\caption{Pseudocolor density plots showing the time evolution of the condensate following a quench of the harmonic trap strength from $\lambda = 1$ to $\lambda = 0.5$. Panel (a) shows the ground-state density profile, while panels (b-d) correspond to the evolution at $t = 100$, 200, and 1600, respectively, for coupling parameters $(k_L,\Omega) = (0.5,1.0)$ and interaction strengths $c_0 = 50$, $c_2 = -2.5$. Columns (i-iii) represent the $j = +1, 0, -1$ components of the spinor condensate. The persistence of the density profiles over long times demonstrates the dynamical stability of the condensate under the trap quench.}
\label{fig8b} 
\end{figure}
In the bottom row, Fig.~\ref{fig08a}(b i) likewise displays only real and positive eigenfrequencies in both the low-lying (denoted by the subscript $ll$) and first-excited (denoted by the subscript $fe$) branches of the eigenspectrum. While the low-lying branch again hosts the phonon mode, the first-excited branch exhibits a pronounced asymmetry in momentum space. The corresponding eigenvector components shown in Fig.~\ref{fig08a}(b ii) display in-phase (density-like) behavior in both branches, satisfying,
\begin{align} \label{eqb:density}
& \vert u_{+1} \vert_{ll}^{2} - \vert u_{-1} \vert_{ll}^{2} = 0, \quad
\vert v_{+1} \vert_{ll}^{2} - \vert v_{-1} \vert_{ll}^{2} = 0, \nonumber \\
& \vert u_{+1} \vert_{fe}^{2} - \vert u_{-1} \vert_{fe}^{2} = 0, \quad
\vert v_{+1} \vert_{fe}^{2} - \vert v_{-1} \vert_{fe}^{2} = 0.
\end{align}
The zeroth component shown in Fig.~\ref{fig08a}(b iii) exhibits a density-like mode for both excitation branches, consistent with previous studies on spin-$1/2$ SO-coupled BECs~\cite{Ravisankar2021}. For the low-lying branch, all eigenvector components converge to the same value at $q_{y} \approx 0$, confirming the phonon character of this mode. In the first-excited branch, the eigenvector components $\vert u_{+1} \vert^2_{fe}$ and $\vert u_{-1} \vert^{2}_{fe}$ remain finite, while $\vert v_{+1} \vert^{2}_{fe}$ and $\vert v_{-1} \vert^{2}_{fe}$ vanish for $q_{y} < -1.08$ and $q_{y} > 1.48$. As $q_{y}$ approaches $-1.08$, the particle-like components $\vert u_{\pm 1} \vert^{2}_{fe}$ begin to decrease, accompanied by the emergence of finite hole-like contributions $\vert v_{\pm 1} \vert^{2}_{fe}$. At $q_{y} \approx -0.21$, all four components attain comparable magnitudes, indicating strong particle–hole mixing. Upon further approaching $q_{y} \approx 0$, the particle-like components reach their maximum values, while the hole-like components again diminish. For $q_{y} > 0$, a similar evolution is observed; however, due to the inherent asymmetry of the excitation spectrum, the recovery of finite $\vert u_{\pm 1} \vert^{2}_{fe}$ and vanishing $\vert v_{\pm 1} \vert^{2}_{fe}$ occurs only near $q{y} \approx 1.48$. Analogous behavior is also present in the zeroth component of the eigenvectors; notably, at $q_{y} \approx 0$, both $\vert u_{0} \vert^{2}_{fe}$ and $\vert v{_0} \vert^{2}_{fe}$ vanish.%

The Bogoliubov spectra and eigenvector analysis presented above reveal clear signatures of mode mixing, roton softening, and particle–hole hybridization, indicating the onset of dynamical instability in specific momentum regimes. Following this, we perform numerical simulations of the coupled Gross–Pitaevskii equations, starting from the mean-field ground states and iniate the dynamics of the condnesate by quenching of the trap.

\paragraph{Numerical Simulation:}
We first obtain the ground state of the condensate for coupling parameters ($k_{L}, \Omega$) = ($0.5, 1.0$) and interaction strengths $c_{0} = 50$, $c_{2} = -2.5$ in presence of the harmonic trap with strength $\lambda = 1.0$. This state exhibits a circularly asymmetric density profile [Figs.~\ref{fig8b}(a i - a iii)]~\cite{Adhikari_2021}. To probe the time evolution of the condensate, we abruptly quench the trap strength from $\lambda = 1.0$ to $\lambda = 0.5$. In the beginning, the condensate changes its shape at $t = 100$ units, as given in Figs.~\ref{fig8b}(b i-b iii). At later times, the density profile begins to recover, and by $t = 200$ units [Fig.~\ref{fig8b}(c i-c iii)] the original shape is largely restored. At $t =1600$ units [Figs.~\ref{fig8b}(d i–d iii)], the condensate fully returns to its initial configuration. This revival indicates that the state is dynamically stable, consistent with the excitation spectrum containing only real eigenfrequencies.%
\begin{figure*}[!htb]
\centering\includegraphics[width=0.99\linewidth]{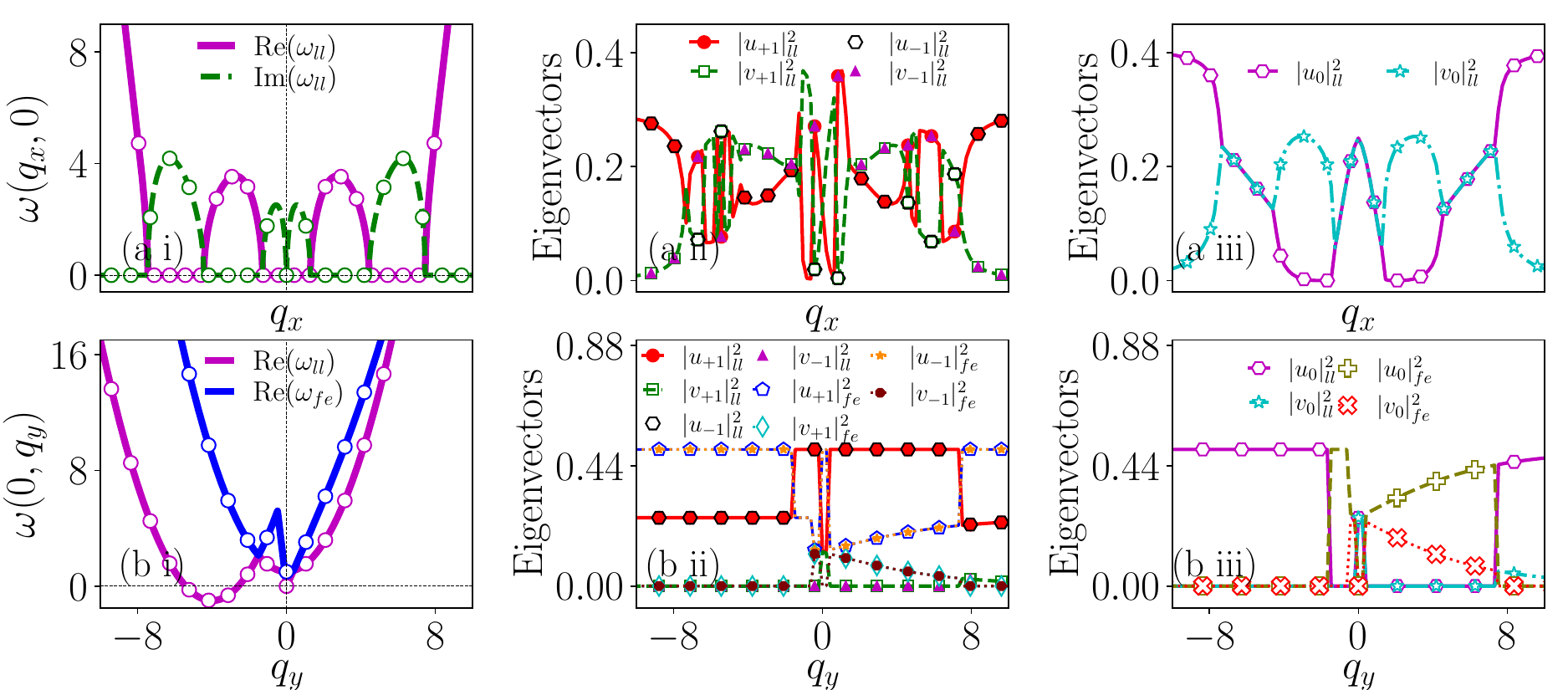}
\caption{Eigenvalue spectra [(a i), (b i)] and corresponding eigenvectors [(a ii, a iii) and (b ii, b iii)] for $k_L = 2.0$ and $\Omega = 3.0$ with interaction strengths $c_0 = 50.0$ and $c_2 = -2.5$, shown along the quasi-momentum directions $q_x$ and $q_y$, respectively. Line styles and symbols follow the conventions of Fig.~\ref{fig08a}. A multiband instability develops along the $q_x$ direction, leading to spin--density--spin mixed modes in the eigenvector components, while the low-lying branch exhibits negative-energy excitations along the $q_y$ direction.}
\label{fig09a} 
\end{figure*}
We report the total energy of the condensate in region I for ($k_{L}, \Omega$) = ($0.5, 1.0$) with interaction strengths $c_{0} = 50$, $c_{2} = -2.5$. The energy starts with a magnitude $E \approx 0.64$ at $t = 0$, which starts to decrease and reaches $E \approx 0.29$ at $t = 370$. Beyond this point, the density of the condensate saturates, showing stable behaviour (not shown here).%

\subsubsection{Region IIa}
\label{sec:5aii}
In region IIa of the stability phase diagram [Fig.~\ref{fig3}(a)], we consider the regimes for which $k_L^2>\Omega$ for which we present here the analysis for two different points ($k_{L}, \Omega$) = ($4.0, 1.0$), and ($4.0, 3.0$). Corresponding to these points, we present the collective excitation spectrum and numerical simulation results separately.%

\paragraph{Collective excitation spectrum:}
In Figure~\ref{fig09a}, we present the collective excitation spectrum for the coupling parameters ($k_{L}$, $\Omega$) = ($4.0, 1.0$) with interaction strengths $c_{0} = 50.0$ and $c_{2} = -2.5$. The top panels display the spectrum along the quasi-momentum direction $q_{x}$, while the bottom panels correspond to the $q_{y}$ direction. In Fig.~\ref{fig09a}(a i), we obtain the multi-band imaginary eigenfrequencies extended over the finite range of quasi-momentum $q_{x} \in [0, 1.30]$ with amplitude $\omega = 4.24$ and $q_{x} \in [4.41, 7.57]$ with amplitude $\omega = 2.55$. The presence of instability bands is symmetric about the quasi-momentum direction. The presence of the imaginary eigenfrequency results in the out-of-phase behaviour (spin-like mode) in eigenvector components [Fig.~\ref{fig09a}(a ii)], which holds the criterion~\cite{Ravisankar2021, Gangwar2025},
\begin{align} \label{eqa:spin}
& \vert u_{+1} \vert_{ll}^{2} - \vert u_{-1} \vert_{ll}^{2} \neq 0, \quad
\vert v_{+1} \vert_{ll}^{2} - \vert v_{-1} \vert_{ll}^{2} \neq 0.
\end{align}
The eigenvector components exhibit the density-like mode in the presence of real eigenfrequencies as discussed in the previous section. Here, in this part, due to the transition of $\text{Re}(\omega) \rightarrow \text{Im}(\omega)$, density-like mode changes to a spin-like mode, and the presence of multiband instabilities leads to density-spin-density or spin-density-spin mixed mode~\cite{Gangwar2024}. The zeroth component of eigenvectors depicts in-phase behaviour (density-like mode) independently [Fig.~\ref{fig09a}(a iii)].

In the bottom row, Fig.~\ref{fig09a}(b i) depicts that the low-lying and first-excited branches of the eigenspectrum contain only real eigenfrequencies, with the low-lying branch developing the negative excitation spectrum. The presence of negative excitation makes the condensate energetically unstable~\cite{Ozawa2013}. Avoided crossings emerge at $q_{y} = -1.50, -0.11$, and $0.43$. Owing to the real eigenfrequencies, the eigenvector components in Fig.~\ref{fig09a}(b ii) exhibit the density-like mode, which follows the criterion given in Eq.~\ref{eqb:density}. At the point of avoided crossings $q_{y} = -4.06, -0.35$ and $0.67$, the eigenvector components undergo a flip~\cite{Abad2013}. The zeroth component of eigenvectors exhibits the density-like mode independently [Fig.~\ref{fig09a}(b iii)]. 
\begin{figure*}[!htb]
\centering\includegraphics[width=0.99\linewidth]{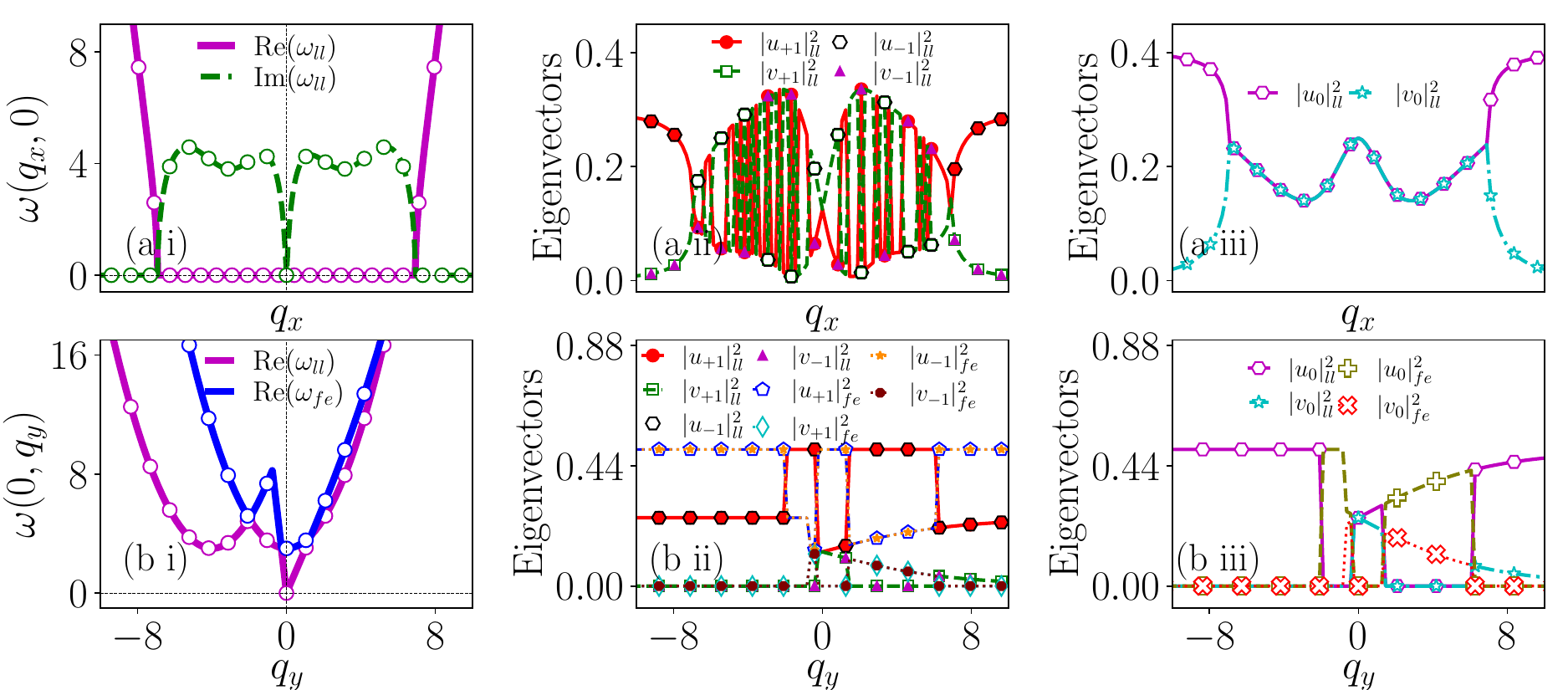}
\caption{Eigenvalue spectra [(a i), (b i)] and corresponding eigenvectors [(a ii, a iii) and (b ii, b iii)] for $k_L = 4.0$ and $\Omega = 3.0$ with interaction strengths $c_0 = 50.0$ and $c_2 = -2.5$, shown along the quasi-momentum directions $q_x$ and $q_y$, respectively. Line styles and symbols follow the conventions of Fig.~\ref{fig08a}. A single-band instability appears in the low-lying branch along the $q_x$ direction, while a roton-maxon-like feature develops in the low-lying branch along the $q_y$ direction.}
\label{fig09b} 
\end{figure*}

In Fig.~\ref{fig09b}, we examine the second point in region IIa, where the coupling parameters are ($k_{L}, \Omega$) = ($4.0, 3.0$) and the interaction strengths are $c_{0} = 50.0$ and $c_{2} = -2.5$. In Fig.~\ref{fig09b}(a i), we observe the presence of an imaginary eigenfrequency in the low-lying branch of the eigenspectrum for a certain range of quasi-momentum direction $q_{x} \in [0, 7.0]$ with maximum amplitude $4.62$. The presence of the imaginary eigenfrequency is symmetric about the quasi-momentum $q_{x}$. There is a transition from $\text{Re}(\omega) \rightarrow$ $\text{Im}(\omega)$. Due to the imaginary eigenfrequencies, as shown in Fig.~\ref{fig09b}(a ii), the eigenvector components exhibit the spin-like mode in the quasi-momentum range $q_{x} \in [0, 7.0]$, which satisfies the criterion given in Eq.~\ref{eqa:spin}. It is symmetric due to the presence of the symmetric instability band in the eigenspectrum, and density-like mode transitions to spin-like mode when $\text{Re}(\omega) \rightarrow$ $\text{Im}(\omega)$. The zeroth component of the eigenvector exhibits density-like mode independently [Fig.~\ref{fig09b}(a iii)]. 

In Fig.~\ref{fig09b}(b i), the low-lying and first-excited branches of the eigenspectrum yield only real, positive eigenfrequencies. Within the low-lying branch, the excitation spectrum exhibits a local minimum at $q_{y} = -4.14$, resembling a roton-like feature. This branch thus contains the characteristic maxon-roton modes. Notably, the structure of the spectrum suggests an asymmetric roton-like feature within the low-lying branch. Avoided crossing between the low-lying and first-excited branch of the eigenspectrum emerges at $q_{y} = -2.08, -0.30$, and $1.39$, respectively. Owing to the real eigenfrequencies, in Fig.~\ref{fig09b}(b ii), the eigenvector components display the density-like mode, consistent with the criterion given in Eq.~\ref{eqb:density}. At the point of avoided crossings $q_{y} = -2.08, -0.30$, and $1.39$, a flip emerges in the eigenvector components. The zeroth component of the eigenvector exhibits the density-like mode independently, and also undergoes a flip at the avoided crossing points [Fig.~\ref{fig09b}(b iii)].%

\begin{figure}[!htp]
\centering\includegraphics[width=0.99\linewidth]{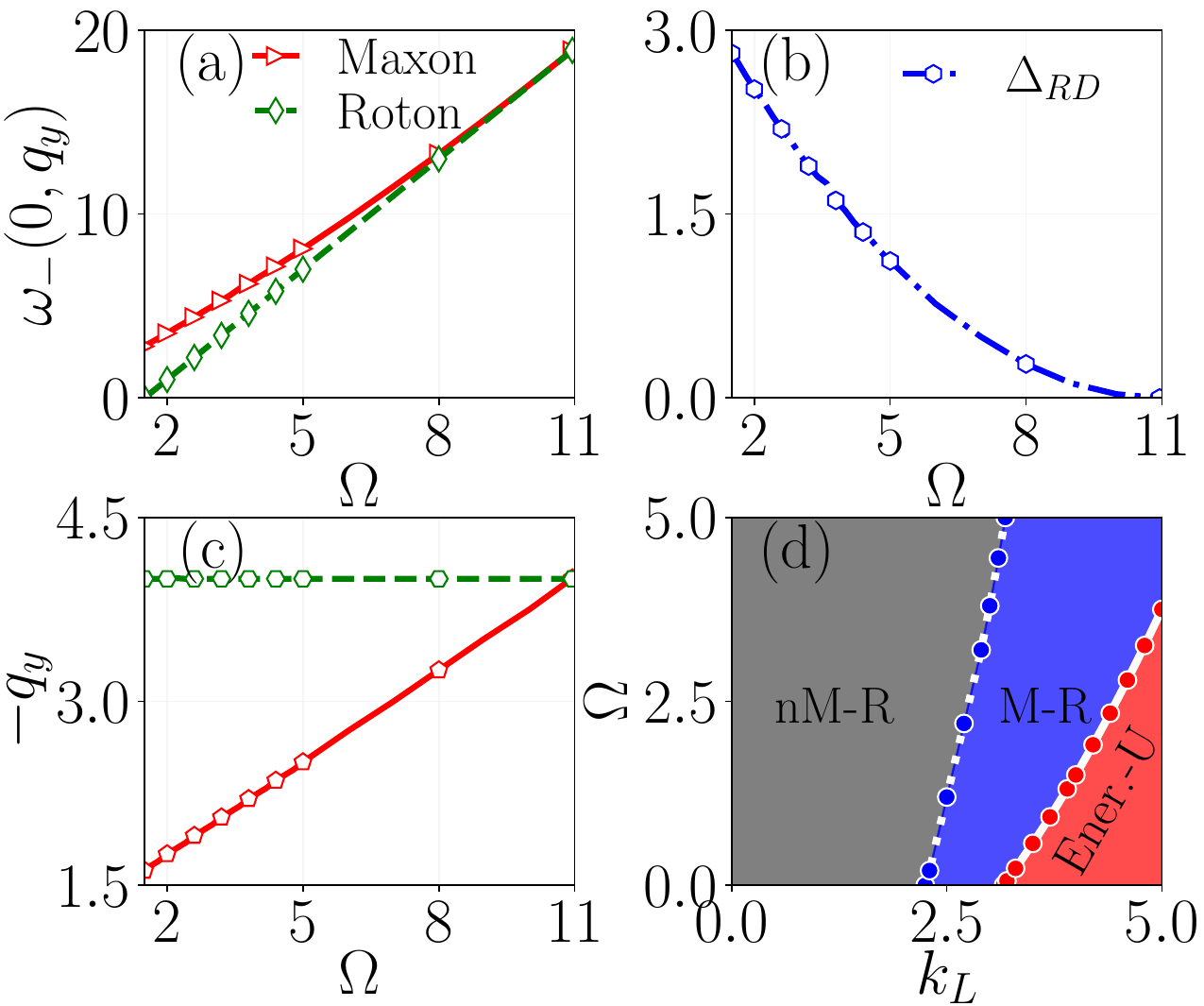}
\caption{Variation of roton-maxon modes for ferromagnetic interaction ($c_{0} = 50$, $c_{2} = -2.5$). Panels (a-c) show the results obtained for $k_{L} = 4.0$ as the Rabi coupling is varied. Panel (a) displays the evolution of the maxon (red open right triangles) and roton (green open diamonds) modes, while panel(b) shows the corresponding decay of the roton depth $\Delta_{RD}$ (blue open hexagons). Panel (c) illustrates quasi-momentum positions of the maxon (red open pentagon) and roton (green open hexagons) modes along the $q_{y}$ direction. Panel (d) presents the phase diagram in the $k_L-\Omega$ plane at $q_{x} = 0.0$, with gray, blue, and red regions indicating the absence of maxon--roton modes (nM--R), the presence of maxon--roton modes (M--R), and energetic instability (Ener.-U), respectively, along the $q_{y}$ direction.}
\label{fig8new} 
\end{figure}

Further, we extend the analysis to characterize the variation of maxon and roton modes by fixing $k_{L} = 4.0$, $c_{0} = 50$, $c_{2} = -2.5$, and varying the Rabi coupling strength. In Fig.~\ref{fig8new}(a), red right open triangles represent the maxon and green open diamonds represent the roton minimum. For $k_{L} = 4.0$, the roton minimum is sustained at $\Omega = 1.50$, with $\omega_{-}(0, q_{y}) = 0.000013$. A further decrease in Rabi coupling leads the excitation branch to negative energy modes. Upon increasing Rabi coupling, the maxon and roton minima approach each other and coincide at $\Omega = 10.95$, where both their magnitudes and positions are equal, and the roton minimum no longer exists beyond this point. Fig.~\ref{fig8new}(b), open blue hexagons indicate the roton depth, defined as the difference between the maxon and roton minima. The Roton depth decays with an increase in Rabi coupling and goes to zero at $\Omega = 10.95$. Fig.~\ref{fig8new}(c) exhibits the positions of maxon (red open pentagons) and roton minimum (green open hexagons). As the Rabi coupling increases, the position of the maxon mode increases, and the roton minimum remains at the same position throughout. At $\Omega = 10.95$, the maxon mode reaches a similar position as the roton minimum and merges with it. Since the maxon and roton modes occur at the negative $q_{y}$, we plot them using $- q_{y}$. 
In Fig.~\ref{fig8new}(d), the $k_{L} -\Omega$ phase plane (along $q_{y}$ direction at $q_{x} = 0.0$) is divided into three distinct regions: the gray (blue) region contains only real, positive eigenfrequencies without (with) maxon-roton modes, separated by a white dashed line with blue dots, and energetically stable. The red region contains negative eigenfrequencies and is energetically unstable. The blue and red region is separated by a solid white line with red dots. Maxon-roton modes occur only in the blue region. Additionally, comparing this with the phase plot Fig.~\ref{fig3}, it is evident that the blue region is a subset of region IIa. The case discussed above with $k_{L} = 4.0$ lies in the blue region and requires a relatively large Rabi coupling, $\Omega = 10.95$, for the maxon and roton to merge. These observations are consistent with previous experimental and theoretical works~\cite{Khamehchi2014,Ha2015,Lyu2020,Lyu2022,Sun2016}.%
\begin{figure*}[!htp]
\centering\includegraphics[width=0.99\linewidth]{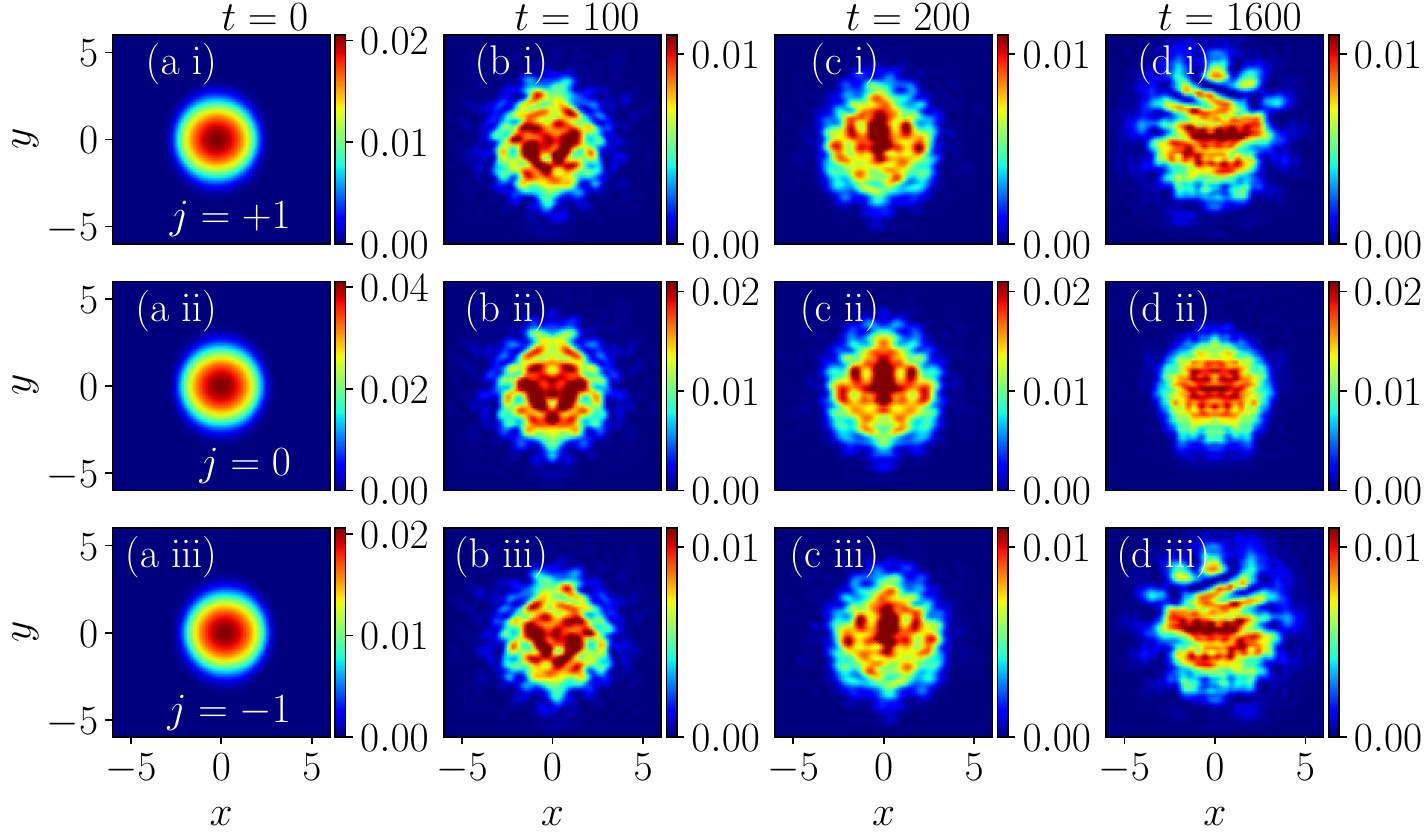}
\caption{Pseudocolor density plots showing the time evolution of the condensate following a quench of the harmonic trap strength from $\lambda = 1$ to $\lambda = 0.5$. (a) Ground state density profile. (b-d) snapshots during time evolution at t = 100, 200, 1600, respectively, for coupling parameters ($k_{L}, \Omega$) = ($4.0, 1.0$), with interaction strengths $c_{0} = 50$, $c_{2} = -2.5$. Each column represents the spinor components, with $j = +1$ (top) , $0$ (middle), and $ -1$ (bottom).}
\label{fig9c} 
\end{figure*}

From the above discussed points, the first point lies in the red region of this phase, which is energetically unstable, and the second point lies in the blue region, energetically stable; however, both points are dynamically unstable due to the presence of imaginary eigenfrequencies.%

\paragraph{Numerical Simulation:}

Similar to the collective excitation discussed in the previous part, we consider two different points to perform the numerical simulation in region IIa. We generate the ground state for coupling parameters $(k_{L}, \Omega)$ = $(4.0, 1.0)$, and $(4.0, 3.0)$ with interaction strengths $c_{0} = 50$, and $c_{2} = -2.5$, and study its dynamics by sudden quenching of trap strength to half of its initial strength.
\begin{figure*}[!htp]
\centering\includegraphics[width=0.99\linewidth]{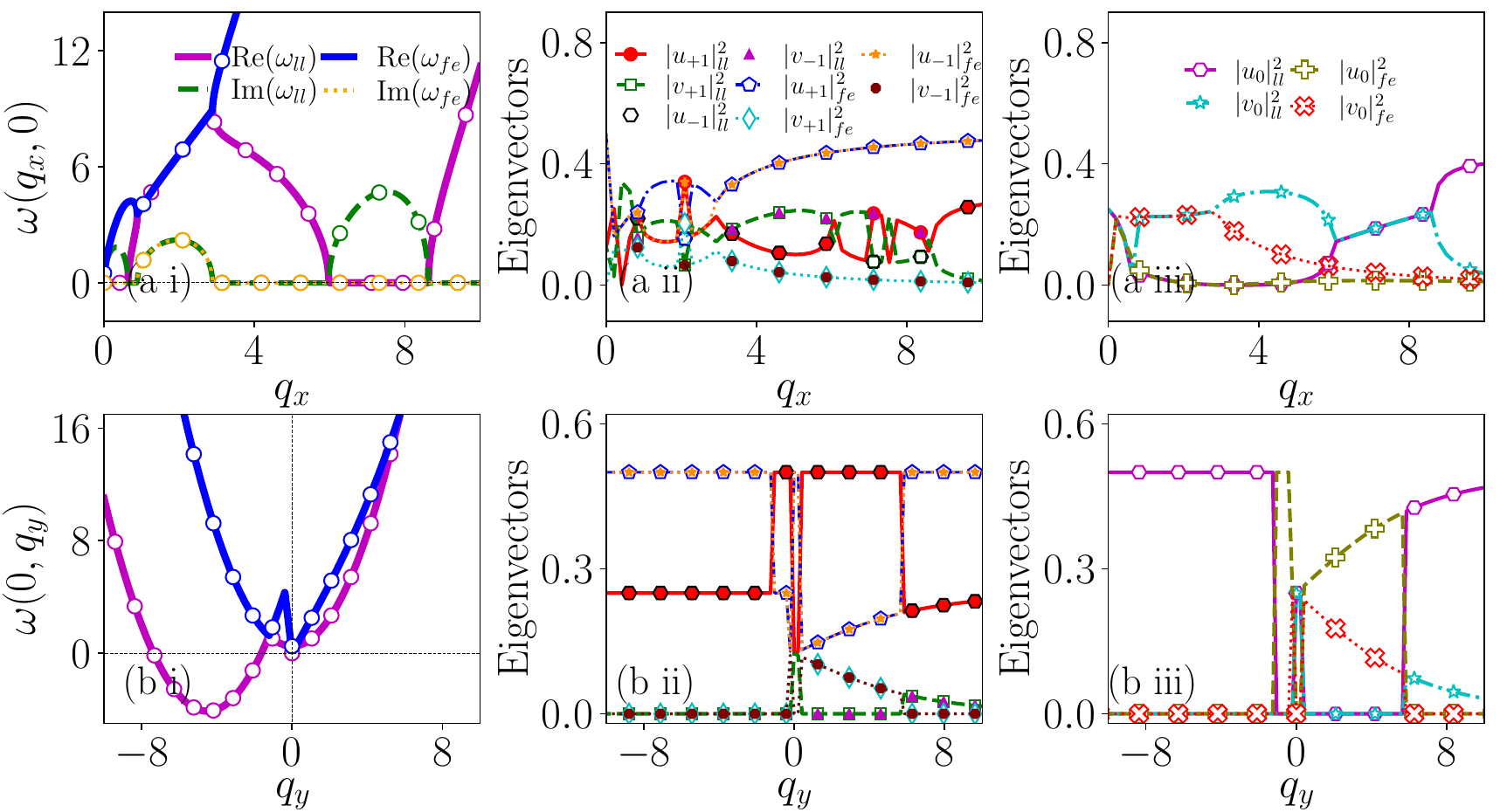}
\caption{Eigenspectra and corresponding eigenvectors for the spinor condensate are shown, with the top (bottom) row corresponding to quasi-momentum along $q_x$ ($q_y$) for $(k_L, \Omega) = (4.5, 0.5)$ and interaction strengths $c_0 = 50.0$, $c_2 = -2.5$. Panels (a~i) and (b~i) display the eigenspectra, with $\mathrm{Re}(\omega)$ shown as solid lines and $|\mathrm{Im}(\omega)|$ as dashed or dotted lines. The eigenvector components for the low-lying and first-excited branches are shown in panels (a~ii, a~iii) and (b~ii, b~iii), respectively. Multi-band instability appears in the low-lying branch, single-band instability in the first-excited branch arises from avoided crossings along $q_x$, and negative eigenfrequencies are observed along $q_y$.}
\label{fig10a} 
\end{figure*}

We first consider the point $(k_{L}, \Omega)$ = $(4.0, 1.0)$ and generate the ground state of the condensate, which is a circularly asymmetric state [Fig.~\ref{fig9c}(a i-a iii)]~\cite{Adhikari_2021}. During time evolution, in Fig.~\ref{fig9c}(b i-b iii), initially the shape of the condensate changes due to a sudden quench of trap strength (as observed in the previous case). At t = 200 units [Fig.~\ref{fig9c}(c i-c iii)], the condensate continues to deform, and at later times $t =1600$ units, it fragments, forming a small domain [Fig.~\ref{fig9c}(d i-d iii)]. The fragmentation of the density profile occurs at the onset of dynamical instability, triggered by the emergence of imaginary frequencies in the excitation spectrum~\cite{Kasamatsu2004, Sadler2006, Mithun2019}.

For the second point in region IIa with coupling parameters ($k_{L}, \Omega$) = ($4.0, 3.0$).  We again obtain a circularly asymmetric state. During dynamics, we quench the trap to half of its original strength during the time evolution. In the time evolution, the shape deformation begins to appear at $t = 100$ units and continues to grow. At $t = 200$, it deforms further, and at a later time $t = 1600$, the fragmentation of density components emerges owing to the presence of an imaginary eigenfrequency in the collective excitation spectrum similar to the previous case (not shown here). 

We calculate the total energy during time evolution of the condensate for the parameter sets discussed above. For the coupling parameters $(k_{L}, \Omega) = (4.0, 1.0)$, the condensate energy begins at $E = -7.30$ at $t = 0$. Following the quench in the trap strength, the system relaxes toward a lower energy configuration, attaining $E = -7.60$ at $t = 250$ units. Beyond this point, the energy remains stable up to $t = 1193.50$ units. Beyond this point, it starts to increase, reaching $E = -5.81$ at $t = 1600$ units (not shown here). For another combination of coupling parameters, $(k_{L}, \Omega) = (4.0, 3.0)$, the condensate starts at $E = -9.29$ at $t = 0$. During the time evolution, the energy decreases to $E = -9.60$ by $t = 275.5$ units and remains nearly constant until $t = 869.5$ units. Thereafter, the energy rises significantly to $E = -0.45$ at $t = 1424.5$ units, before relaxing again to $E = -1.23$ at $t = 1600$ units (not shown here). The observed temporal modulation of the total energy in both cases reflects the dynamical instability of the condensate following the quench~\cite{Ravisankar2021}.%

\subsubsection{Region IIb}
\label{sec:5aiii}
In the stability region IIb, we consider the point with coupling parameters ($k_{L}, \Omega$) = (4.5, 0.5). The collective excitation spectrum and numerical simulation results for this case are presented in the following section.
\begin{figure*}[!htp]
\centering\includegraphics[width=0.99\linewidth]{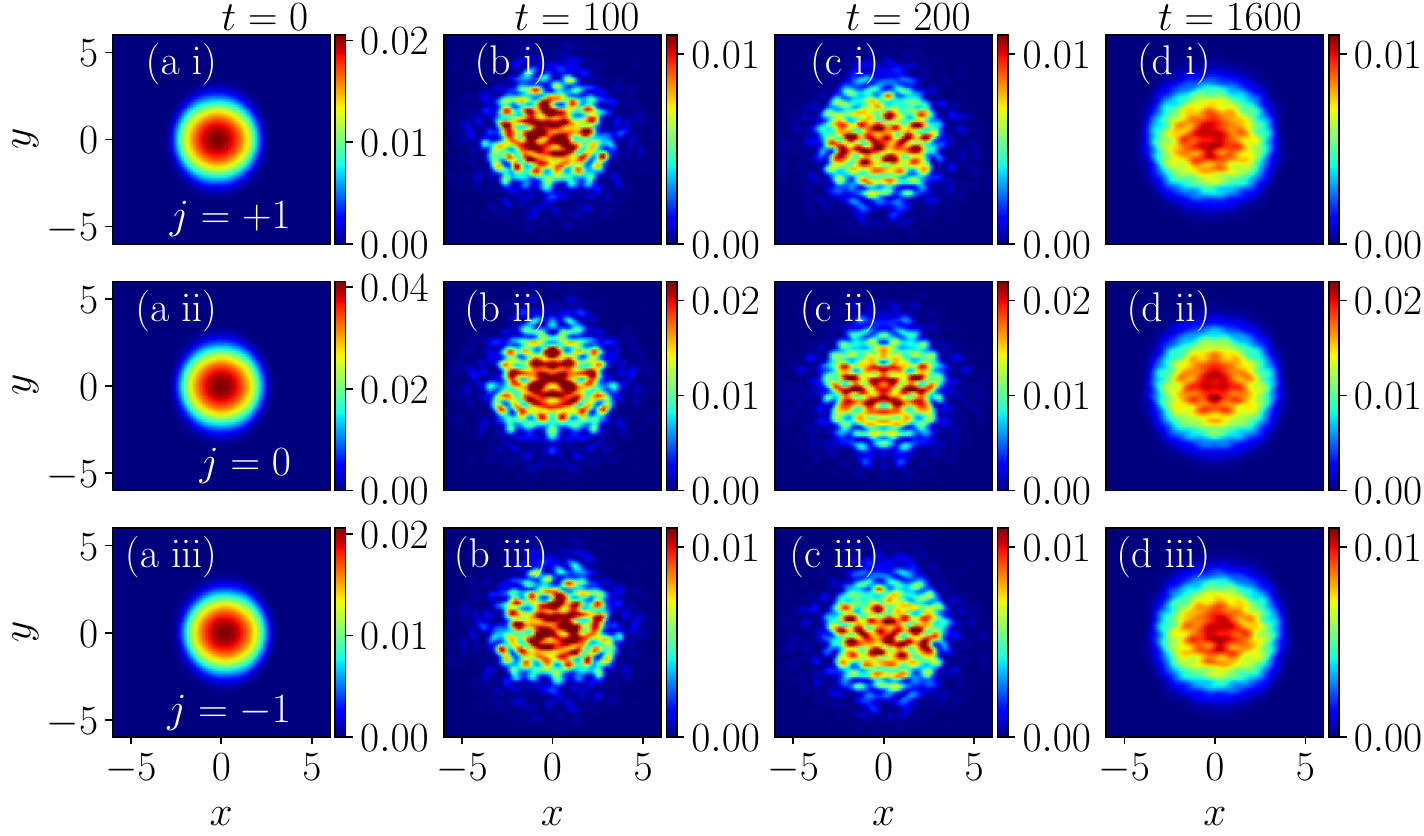}
\caption{Pseudocolor plots in panel represent (a) ground state density and (b-d) density profile during time evolution at times $t = 100, 200, 1600$, respectively, for coupling parameters ($k_{L}, \Omega$) = ($4.5, 0.5$) with interaction strengths $c_{0} = 50$, $c_{2} = -2.5$. Each column (i–iii) corresponds to the density of the spinor components $\vert \psi_{+1} \vert^{2}$, $\vert \psi_{0} \vert^{2}$, and $\vert \psi_{-1} \vert^{2}$, respectively. During the time evolution, the trap strength is suddenly reduced to half of its original value.}
\label{fig10b} 
\end{figure*}
\paragraph{Collective excitation spectrum:} In Fig.~\ref{fig10a}, we present the collective excitation spectrum for spin–orbit and Rabi coupling parameters ($k_{L}, \Omega$) = ($4.5, 0.5$), with interaction strengths $c_{0} = 50.0$ and $c_{2} = -2.5$. The spectrum is shown along the quasi-momentum directions $q_{x}$ (top row) and $q_{y}$ (bottom row). In Fig.~\ref{fig10a}(a i), the low-lying branch of the eigenspectrum exhibits three bands of imaginary eigenfrequencies. The first band extends over $q_{x} \in [0, 0.61]$ with an amplitude of $1.90$, the second over $q_{x} \in [0.87, 2.87]$ with an amplitude of $2.20$, and the third over $q_{x} \in [5.98, 8.65]$ with an amplitude of $4.83$. In contrast, the first excited branch displays only a single band of imaginary eigenfrequency. This arises from an avoided crossing between the low-lying and first excited branches of the eigenspectrum. The band extends over $q_{x} \in [0.87, 2.87]$ with an amplitude of $2.20$, coinciding with the second band of the low-lying branch. Since this avoided crossing produces imaginary eigenfrequencies, it corresponds to an unstable avoided crossing~\cite{Bernier2014, Gangwar2024}. Owing to the presence of the imaginary eigenfrequencies, the eigenvector components in the low-lying branch and in the first-excited branch in Fig.~\ref{fig10a}(a ii) exhibit the out-of-phase behaviour (spin-like mode) following the criterion,
\begin{align} \label{eqb:spin}
& \vert u_{+1} \vert_{ll}^{2} - \vert u_{-1} \vert_{ll}^{2} \neq 0, \quad
\vert v_{+1} \vert_{ll}^{2} - \vert v_{-1} \vert_{ll}^{2} \neq 0, \nonumber \\
& \vert u_{+1} \vert_{fe}^{2} - \vert u_{-1} \vert_{fe}^{2} \neq 0, \quad
\vert v_{+1} \vert_{fe}^{2} - \vert v_{-1} \vert_{fe}^{2} \neq 0.
\end{align}
At the point of unstable avoided crossing, the eigenvector components of the low-lying and first excited branches are out of phase with each other for $q_{x} \\in [0.87, 2.87]$. The low-lying branch shows multiband instability, resulting in spin-density-spin mixed modes in the eigenvector components. The zeroth component of the eigenvectors independently exhibits the density-like mode [Fig.~\ref{fig10a}(a iii)].

In the bottom row along $q_{y}$ direction, Fig.~\ref{fig10a}(b i) shows only real eigenfrequencies in both the low-lying and first-excited branches of the eigenspectrum. An avoided crossing occurs between these branches at $q_{y} = -1.22$. We also observe a negative eigenfrequency in the low-lying branch. The real eigenfrequencies lead the eigenvector components in Fig.~\ref{fig10a}(b ii) to display in-phase behavior (density-like mode), as defined in Eq.~\ref{eqb:density}. At the avoided crossing, the eigenvector components flip~\cite{Abad2013}. The zeroth component independently displays in-phase behavior [Fig.~\ref{fig10a}(b iii)]. The presence of imaginary and negative eigenfrequencies causes dynamic and energetic instability throughout~\cite{Ozawa2013}.

\paragraph{Numerical Simulation:}

In this part of the study, we present numerical simulations in region IIb with coupling parameters $k_{L} = 4.5$ and $\Omega = 0.5$, and interaction strengths $c_{0} = 50$ and $c_{2} = -2.5$. The initial ground state obtained is circularly asymmetric [Fig.~\ref{fig10b}(a i-a iii)]~\cite{Adhikari_2021}. To study the condensate dynamics, the trap strength is suddenly reduced to half its original value. As shown in Fig.~\ref{fig10b}(b i-b iii), the condensate deforms and changes shape at $t = 100$ units. By $t = 200$ units, it further fragments, though its amplitude is lower than before [Fig.~\ref{fig10b}(c i-c iii)]. At later times, the amplitude continues to decrease, and the condensate evolves into a stripe-like structure across all components [Fig.~\ref{fig10b}(d i-d iii)], which validates the presence of dynamical instability of the condensate in the region~\cite{Sadler2006, Mithun2019}.%

\begin{figure}[!htp]
\centering\includegraphics[width=0.99\linewidth]{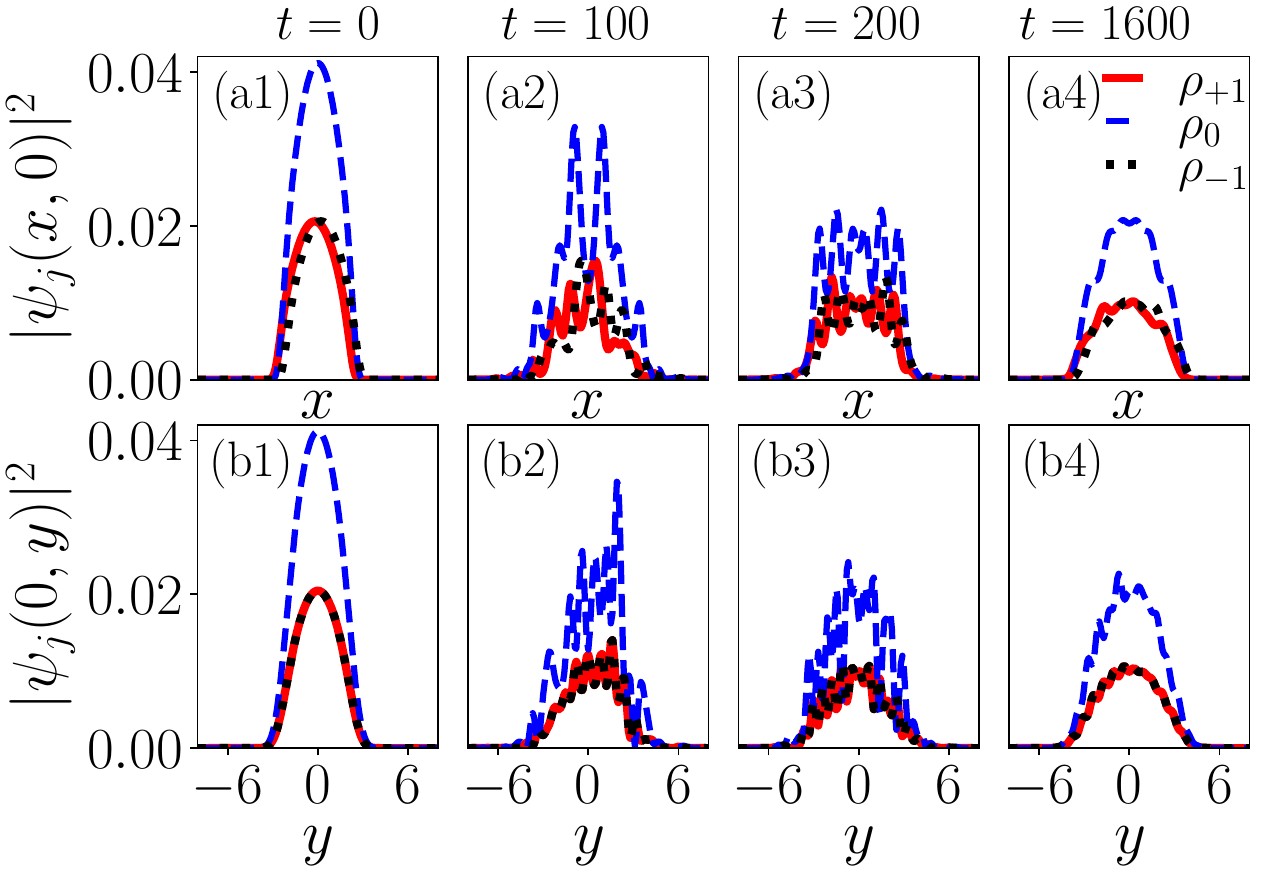}
\caption{(a) ground state density profile and (b-d) density profiles at $t = 100, 200, 1600$, respectively. The top row depicts densities along the spatial direction $x$ at $y = 0$, while the bottom row displays the densities along the spatial direction $y$ at $x = 0$. The coupling parameters and interaction strengths are similar as in Fig.~\ref{fig10b}. The solid red line represents the $\vert \psi_{+1} \vert^{2}$, the dashed blue line represents $\vert \psi_{0} \vert^{2}$, and dotted black line represents $\vert \psi_{-1} \vert^{2}$.}
\label{fig10c} 
\end{figure}

In Fig.~\ref{fig10c}, we present line plots corresponding to the density profiles shown in Fig.~\ref{fig10b}. The top row displays densities along the $x$-direction at $y = 0$, while the bottom row exhibits densities along the $y$-direction at $x = 0$. Along the $x$-direction, Fig.~\ref{fig10c}(a1) depicts the ground state of the condensate with a single-peaked density for all components, with $\rho_{\pm 1}$ slightly shifted from the trap center. At $t = 100$ units, owing to the onset of dynamical instability, the condensate shape evolves, and multiple peaks appear across the density profiles [Fig.~\ref{fig10c}(a2)]. By $t = 200$ units, the densities fragment further [Fig.~\ref{fig10c}(a3)], and at $t = 1600$ units, the condensate develops a stripe-like structure [Fig.~\ref{fig10c}(a4)]. Along the $y$-direction, in Fig.~\ref{fig10c}(b1), we report the ground state with the single peak density profile across all components. During dynamics, owing to the onset of dynamical instability, the shape of the condensate changes, and it fragments into several small domains [Fig.~\ref{fig10c}(b2, b3)]. At a later time, at $t = 1600$ units, condensate amplitude diminishes further, showing the stripe-like density profile across components [Fig.~\ref{fig10c}(b4)]~\cite{Kasamatsu2004}.%

\begin{figure*}[!htp]
\centering\includegraphics[width=0.99\linewidth]{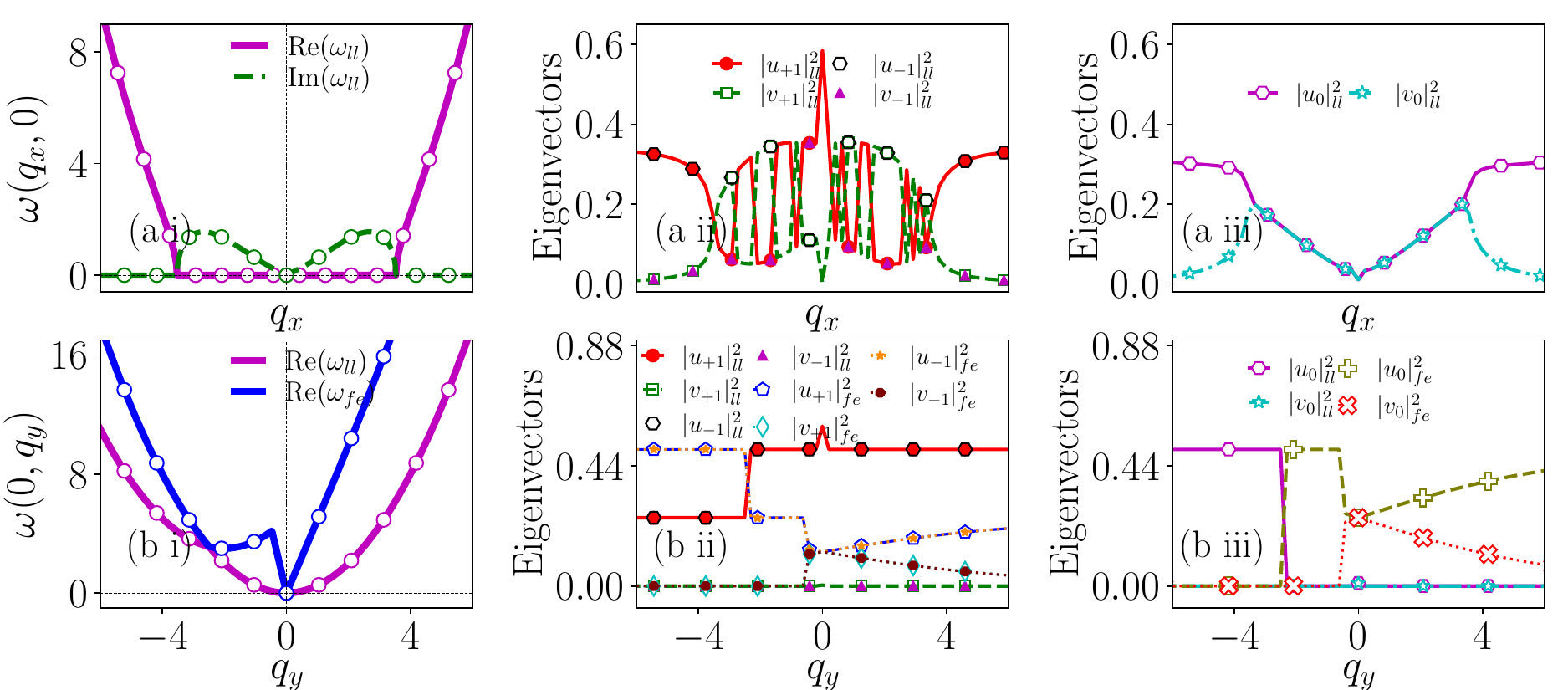}
\caption{Collective excitation spectrum (a i) and (b i) with the corresponding eigenvectors (a ii, a iii) and (b ii, b iii) for $k_{L} = 2.0$, $\Omega \sim 0.0$, using interaction strengths $c_{0} = 50.0$ and $c_{2} = -2.5$ along with quasi-momentum directions $q_{x}$ and $q_{y}$, respectively. The line styles and symbols in the top and bottom rows follow the same convention as in Fig.~\ref{fig08a}.}
\label{fig4a1} 
\end{figure*}
Corresponding to the above point in region IIb for the coupling parameters ($k_{L}, \Omega$) = ($4.5, 0.5$), we report the total energy of the condensate during dynamics. The energy of the condensate starts at $E = -8.93$ at $t = 0$ and decreases during the dynamics, reaching $E = -9.24$ at $t = 792$ units and $E = -9.25$ at $t = 998$ units. By $t = 1600$, it further reduces to $E = -9.26$. The energy decreases very slowly and becomes almost constant at later times [not shown here].%


\subsection{Collective excitation spectrum in absence of Rabi coupling ($\Omega$): Region III}
\label{sec:5b}
In the previous section, we discussed the collective excitation spectrum with both SO and Rabi coupling, covering regions I, IIa, and IIb of the stability phase diagram [Fig.~\ref{fig3}]. Now, we focus on the case without Rabi coupling ($\Omega \sim 0$). We refer to this regime as region III of the stability phase diagram and present both the collective excitation spectrum and the corresponding numerical simulation results.%

In region III of the stability phase diagram [Fig.~\ref{fig3}], we examine two representative points with coupling parameters ($k_{L}, \Omega$) = ($2.0, 0.0$) and ($4.0, 0.0$) in which we have $k_L^\Omega$, but we consider $\Omega\sim 0$. To investigate the behavior of the condensate in this regime, we analyze the collective excitation spectrum together with numerical simulations. 

\paragraph{Collective excitation spectrum:} In Fig.~\ref{fig4a1}, we present the eigenvalue spectrum for coupling parameters ($k_{L}, \Omega$) = ($2.0, 0.0$) and interaction strengths $c_{0} = 50.0$, $c_{2} = -2.5$. The top row presents the eigenspectrum along the quasi-momentum direction $q_{x}$, while the bottom row corresponds to the $q_{y}$ direction of quasi-momentum. In Fig.~\ref{fig4a1}(a i), we obtain the imaginary eigenfrequency in the spectrum for a finite range of quasi-momentum direction $q_{x} \in [0, 3.55]$ with band amplitude of about $1.56$. The presence of the imaginary eigenfrequency is symmetric about $q_{x}$. The corresponding eigenvectors in Fig.~\ref{fig4a1}(a ii) display the spin-like mode in a certain quasi-momentum range $q_{x} \in [0, 3.55]$, consistent with the criterion in Eq.~\ref{eqa:spin}, owing to the presence of the imaginary eigenfrequencies. A kink-like feature emerges in the $\vert u_{+1}\vert^{2}_{ll}$ and $\vert v_{+1}\vert^{2}_{ll}$ at $q_{x} \approx 0$, which was absent in the presence of the Rabi coupling strength. The zeroth component of the eigenvectors exhibits the density-like mode independently [Fig.~\ref{fig4a1}(a iii)]. In the bottom row, in Fig.~\ref{fig4a1}(b i), we obtain only real, positive eigenfrequencies in the eigenspectrum in the low-lying and first-excited branch. An avoided crossing occurs between the low-lying branch and first-excited branch at $q_{y} = -2.5$. Owing to the real eigenfrequencies, in Fig.~\ref{fig4a1}(b ii), the eigenvector exhibits the density-like mode, which holds the criterion Eq.~\ref {eqb:density}.  At avoided crossing, the eigenvector undergoes a flip at $q_{y} = -2.5$~\cite{Abad2013}. A similar kink-like structure appears in $\vert u_{+1}\vert^{2}_{ll}$ and $\vert v_{+1}\vert^{2}_{ll}$ at $q_{x} \approx 0$. The zeroth component of the eigenvectors depicts the density-like mode independently [Fig.~\ref{fig4a1}(b iii)].%

\begin{figure*}[!htp]
\centering\includegraphics[width=0.99\linewidth]{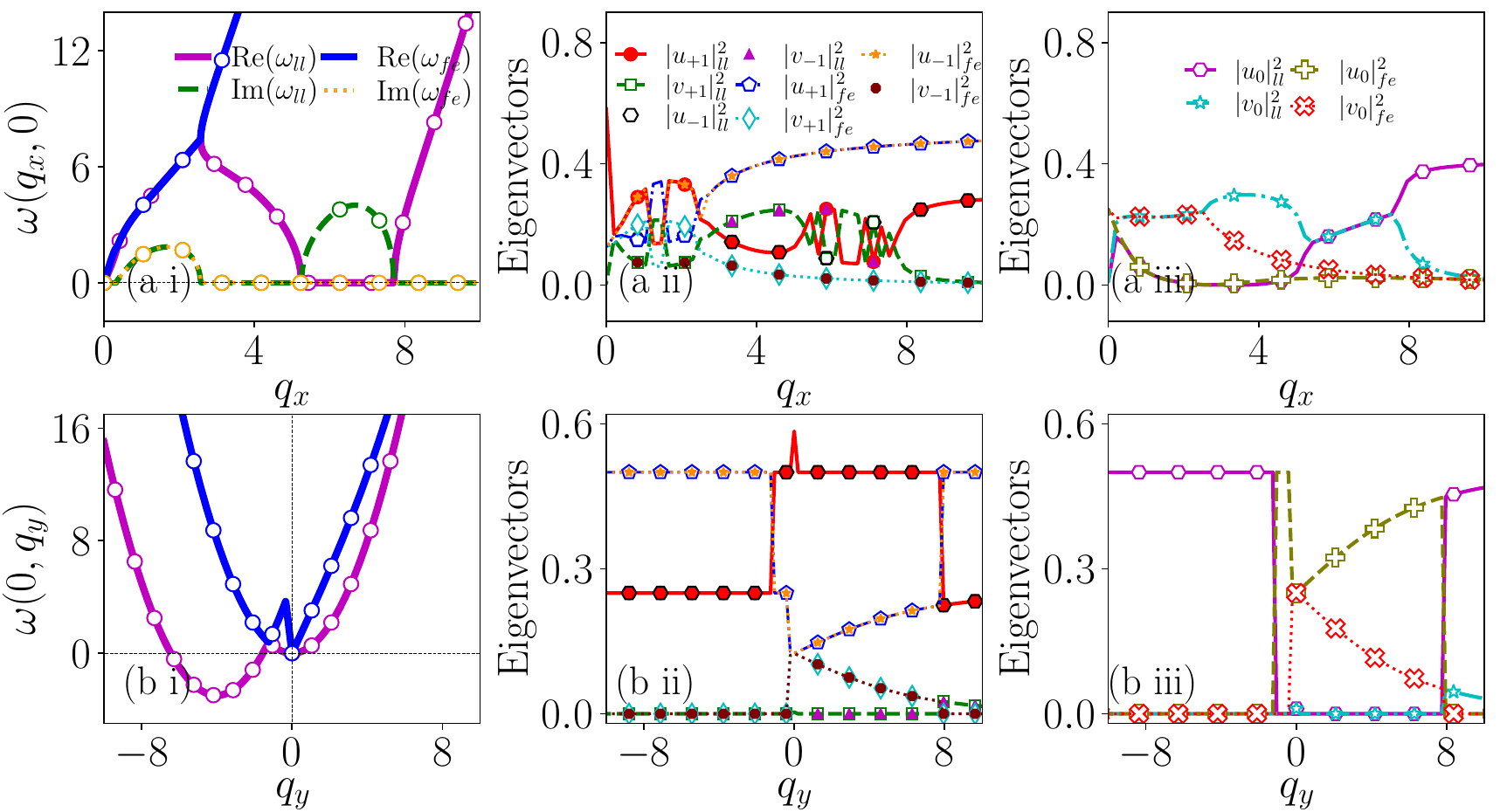}
\caption{(a i, b i) Eigenvalue spectrum and (a ii, a iii, b ii, b iii) corresponding eigenvectors for $k_{L} = 4.0$ in the static limit ($\Omega = 0$), with interaction strengths $c_{0} = 50.0$ and $c_{2} = -2.5$. The spectra and eigenvectors are plotted as functions of the quasi-momenta $q_{x}$ and $q_{y}$, respectively. The line styles and symbols in the top and bottom rows follow the same convention as in Fig.~\ref{fig10a}.}
\label{fig4a3} 
\end{figure*}

In Fig.~\ref{fig4a3}, we report the collective excitation spectrum for the coupling parameters ($k_{L}, \Omega$) = ($4.0, 0.0$) with interaction strengths $c_{0} = 50.0$ and $c_{2} = -2.5$. The top panel displays the eigenvalue spectrum as a function of the quasi-momentum $q_{x}$, while the bottom panel exhibits the spectrum along the $q_{y}$ direction. In Fig.~\ref{fig4a3}(a i), the low-lying branch of the eigenspectrum exhibits the presence of multi-band imaginary eigenfrequencies for $q_{x} \in [0, 3.30]$ with band amplitude $2.93$, and $q_{x} \in [6.26, 8.74]$ with band amplitude $4.74$. In the first-excited branch of the eigenspectrum, we obtain the presence of a single band instability for quasi-momentum range $q_{x} \in [0, 3.30]$ with band amplitude $2.93$, which emerges as a result of unstable avoided crossing between the low-lying and first-excited branches of the eigenspectrum~\cite{Bernier2014, Gangwar2024}. The location and amplitude are similar to those of the first band on the low-lying branch. The corresponding eigenvector components in Fig.~\ref{fig4a3}(a ii) display the spin-like mode due to the presence of imaginary eigenfrequencies, following the criterion given in Eq.~\ref{eqb:spin}. At the point of unstable avoided crossing for $q_{x} \in [0, 3.30]$, the eigenvector components of low-lying and first-excited branches are out of phase among themselves as well as out of phase with each other. Similar to the previous case, a kink-like structure emerges in $\vert u_{+1}\vert^{2}_{ll}$ and $\vert v_{+1}\vert^{2}_{ll}$ at $q_{x} \approx 0$. The zeroth component of the low-lying and first-excited branches displays the density-like mode independently [Fig.~\ref{fig4a3}(a iii)]. In the bottom row, Fig.~\ref{fig4a3}(b i) presents the eigenspectrum along the quasi-momentum direction $q_{y}$. The spectrum exhibits only real eigenfrequencies for both the low-lying and first-excited branches. An avoided crossing appears between these two branches at $q_{y} = -1.09$, and the low-lying branch takes on negative eigenvalues. The presence of negative excitations leads to energetic instability of the condensate~\cite{Ozawa2013}. The corresponding eigenvector components in Fig.~\ref{fig4a3}(b ii) display the density-like mode for low-lying and the first-excited branches. A kink-like structure emerges in $\vert u_{+1}\vert^{2}_{ll}$ and $\vert v_{+1}\vert^{2}_{ll}$ at $q_{x} \approx 0$. At the point of the avoided $q_{y} = -1.09$, eigenvector components undergo a flip for both branches. The zeroth component of the eigenvectors exhibits the density-like mode independently [Fig.~\ref{fig4a3}(b iii)]. The region is dynamically unstable throughout; however, both energetically and dynamically unstable for higher SO coupling strength [Fig.~\ref{fig8new}(d)].%

\begin{figure*}[!htp]
\centering\includegraphics[width=0.99\linewidth]{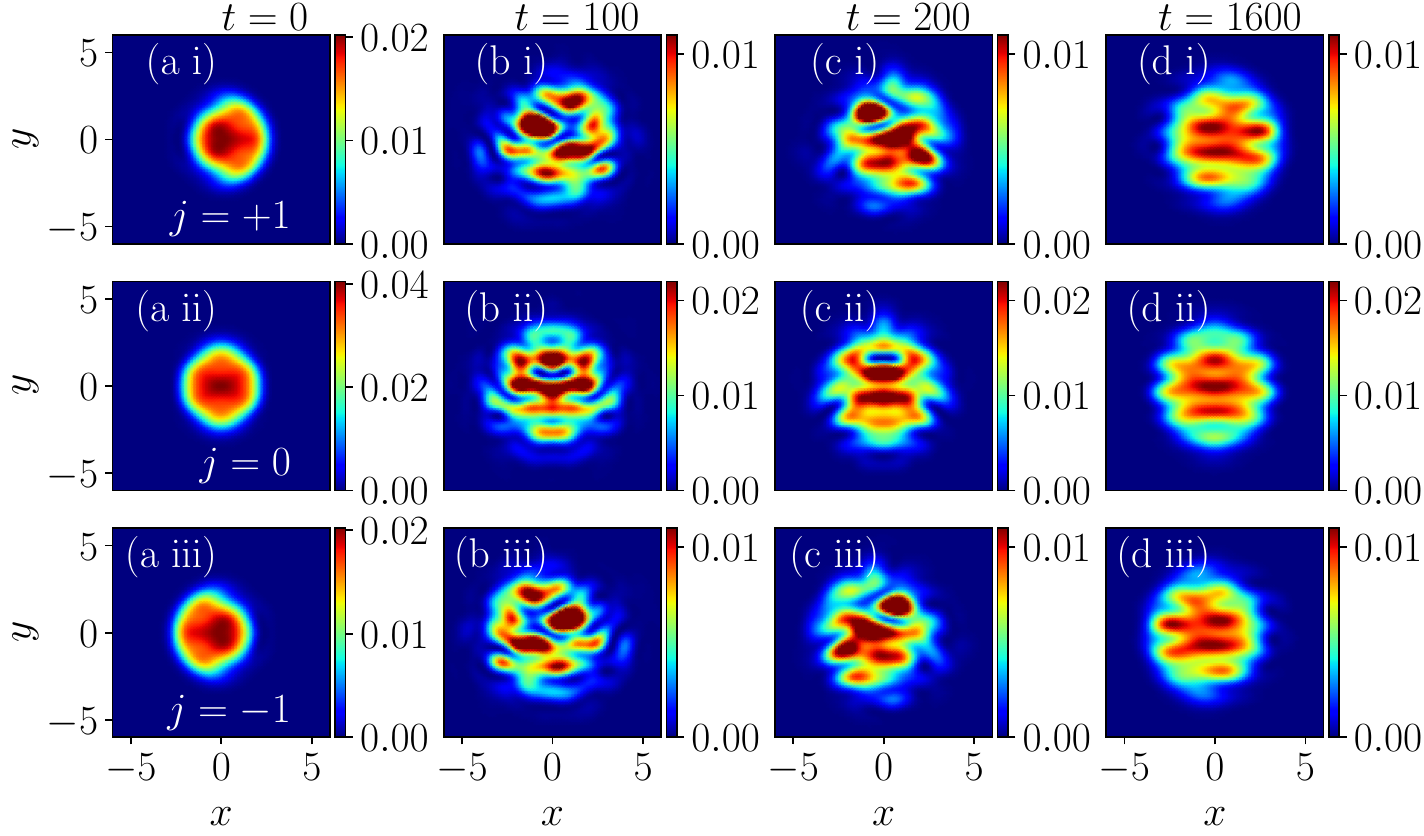}
\caption{Pseudocolor plots of the ground state and time-evolved densities. (a) The circularly asymmetric density profile of the ground state. (b–d) The density evolution at times $t = 100, 200,$ and $1600$, respectively. Columns (i)–(iii) correspond to the spinor components $j = +1, 0,$ and $-1$, respectively. Parameters: lattice coupling $(k_{L}, \Omega) = (2.0, 0.0)$ and interaction strengths $c_{0} = 50$, $c_{2} = -2.5$.}
\label{fig4a4} 
\end{figure*}

\paragraph{Numerical Simulation:} In this section, we present numerical simulations for region III, focusing on two parameter sets: ($k_{L}, \Omega$) = ($2.0, 0.0$) and ($4.0, 0.0$), with interaction strengths $c_{0} = 50$ and $c_{2} = -2.5$. First, we report the numerical simulation for $k_{L} = 2.0$ in the absence of Rabi coupling ($\Omega = 0$). In Fig.~\ref{fig4a4}(a i-a iii), we obtain the ground state density profile, which is a circularly highly asymmetric state. The stripe phase also exists for the above-mentioned SO coupling strength in the absence of Rabi coupling strength, which is an excited state of the condensate~\cite{Adhikari_2021, Adhikari2021127042}. To study the quench dynamics, we abruptly reduce the trap strength to half its initial value. As the time evolves, the condensate shape deforms and breaks into several small domains across components at $t = 100$ and $200$ units [Figs.~\ref{fig4a4}(b i–b iii, c i–c iii)]. At a later time ($t = 1600$ units), as shown in Fig.~\ref{fig4a4}(d i–d iii), the amplitudes of the densities $\rho_{\pm 1}$ decrease, and stripe-like patterns begin to appear across the components~\cite{Sadler2006, Mithun2019}. 


Next, we examine the second point in region III with coupling parameters ($k_{L}, \Omega$) = ($4.0, 0.0$). The initially obtained ground-state density profile is circularly asymmetric. Similar to the above case, the stripe phase also exists for SO coupling mentioned here in the absence of Rabi coupling, which is an excited state as well~\cite{Adhikari_2021,Adhikari2021127042}. As before, we abruptly change the trap strength by half to study the dynamics of the condensate. Since the eigenspectrum in Fig.~\ref{fig4a3}(a i) contains the imaginary eigenfrequency, the condensate fragments into several smaller domains during evolution. By $t = 100$ units, fragmentation is already visible across all components, and it becomes more pronounced at later times ($t = 200$ units). At a later time, $t = 1600$ units, the density of the zero-component is reduced, and the system exhibits a strongly fragmented structure across components (not shown here).

We now report the total energy of the condensate during time evolution in region III for above mentioned points $(k_{L}, \Omega)$ = ($2.0, 0.0$) and ($4.0, 0.0$). Initially, we consider $k_{L} = 2.0, \Omega = 0.0$. The energy starts with $E = -0.31$ at $t = 0$ units. During time evolution, following the quench of the trap strength, the energy of the condensate relaxes to a lower value, reaching $E = -0.64$ at $t = 1600$ units (not shown here). The total energy of the condensate behaves differently at another point in the region, where we consider ($k_{L}, \Omega$) = ($4.0, 0.0$). The energy of the condensates starts with a value $E = -6.4$ at $t = 0$ units. It starts to decrease similarly to the previous point and reaches $E = -6.6$ at $t = 253$ units, remaining flat until $t = 1165.75$ units. It rises again and reaches a higher value $E = -3.7$ at $t = 1600$ units (not shown here).%


\section{Collective Excitations for Anti-ferromagnetic interactions}
\label{sec:stabphaseaferro}
So far, we have focused on the stability analysis in the presence of ferromagnetic interactions across the $k_L - \Omega$ plane, where we identified stable region I and unstable region II, subdivided into IIa and IIb. We also designated the line corresponding to zero Rabi coupling ($\Omega = 0$) as region III, which remains unstable throughout. For $\Omega\sim 0$ we have recently reported the emergence of the superstripe in one-dimensional SO coupled spin-1 BECs with anti-ferromagnetic interaction~\cite{gangwar2025emergence}. Here we extend the work for 2D Spin-1 BEs, for the case of antiferromagnetic interactions, we now examine the scenario with zero Rabi coupling for two different SO coupling strengths, $k_L = 2.0$ and $4.0$, with interaction strengths $c_0 = 50$ and $c_2 = 2.5$. These parameter choices correspond to antiferromagnetic interactions, where the ground state at $\Omega = 0$ forms a superstripe (supersolid-like) phase~\cite{Adhikari_2021}. Since a finite Rabi coupling tends to suppress this phase, restricting our analysis to $\Omega = 0$ allows us to emphasize the intrinsic behavior of the system without Rabi-induced mixing.
\subsubsection*{Case (i): $k_{L} = 2.0, \Omega = 0.0$}
We begin by considering the antiferromagnetic regime with interaction strengths $c_{0} = 50$, $c_{2}= 2.5$. We first focus on the SO coupling strength $k_{L} = 2.0$ in the absence of the Rabi coupling ($\Omega = 0.0$), and present the corresponding collective excitation spectrum and condensate dynamics.%

\begin{figure}[!htb] 
\begin{centering}
\centering\includegraphics[width=0.95\linewidth]{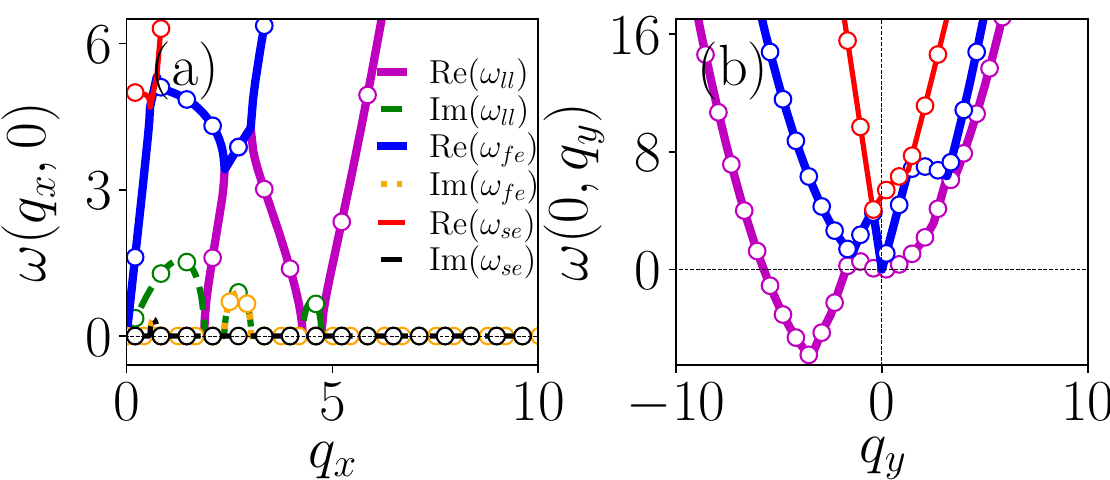}
\caption{Collective excitation spectrum as a function of (a) quasi-momentum \(q_{x}\) and (b) quasi-momentum \(q_{y}\). Parameters are lattice coupling strength \(k_{L}=2.0\), Rabi frequency \(\Omega=0.0\), and interaction strengths \(c_{0}=50\), \(c_{2}=2.5\). Solid magenta, solid blue, and thin red lines represent \(\text{Re}(\omega_{ll})\), \(\text{Re}(\omega_{fe})\), and \(\text{Re}(\omega_{se})\), respectively; dashed green, dotted orange, and dash-dotted black lines show the corresponding magnitude of the imaginary part \(|\text{Im}(\omega)|\). A multiband instability with a roton minimum appears along \(q_x\), while negative excitation frequencies are observed along \(q_y\).}
\label{figaferroeval}
\end{centering}
\end{figure}%

\begin{figure*}[!htb] 
\begin{centering}
\centering\includegraphics[width=0.99\linewidth]{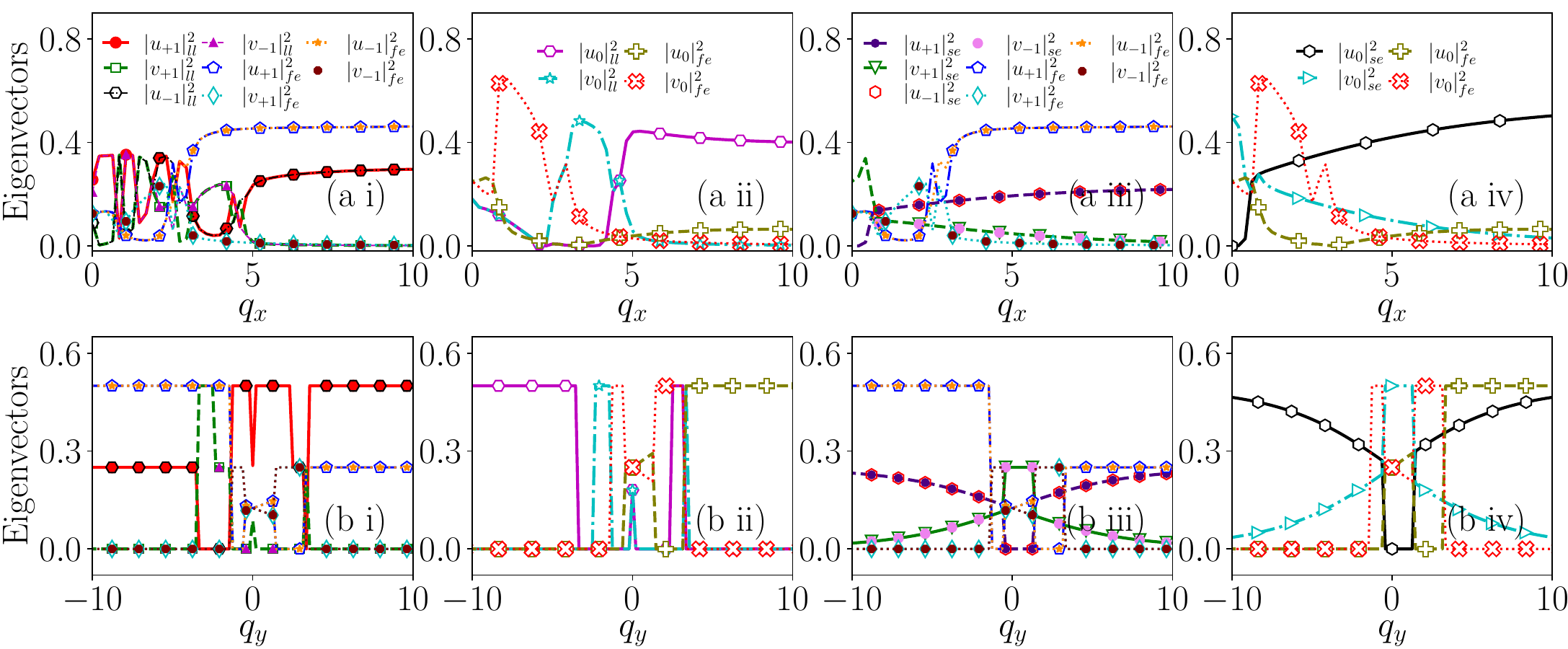}
\caption{Eigenvector components for the excitation branches along the quasi-momentum directions (a\,i--a\,iv) $q_x$ and (b\,i--b\,iv) $q_y$, corresponding to the eigenvalues in Fig.~\ref{figaferroeval}. For each branch, squared amplitudes $|u_{m}|^2$ and $|v_{m}|^2$ (with $m = 0, \pm1$) are shown using distinct markers. Low-lying branch: $|u_{+1}|^2_{ll}$ (red open circles), $|u_{-1}|^2_{ll}$ (black diamonds), $|u_{0}|^2_{ll}$ (magenta open hexagons), $|v_{0}|^2_{ll}$ (cyan open stars), $|v_{+1}|^2_{ll}$ (green open squares), $|v_{-1}|^2_{ll}$ (magenta triangles). First-excited branch: $|u_{+1}|^2_{fe}$ (blue open pentagons), $|u_{-1}|^2_{fe}$ (orange stars), $|u_{0}|^2_{fe}$ (olive open pluses), $|v_{0}|^2_{fe}$ (red open crosses, dotted line), $|v_{+1}|^2_{fe}$ (cyan open diamonds), $|v_{-1}|^2_{fe}$ (maroon dots). Second-excited branch: $|u_{+1}|^2_{se}$ (indigo dots), $|u_{-1}|^2_{se}$ (red hexagons), $|u_{0}|^2_{se}$ (black hexagons), $|v_{0}|^2_{se}$ (cyan left-pointing triangles), $|v_{+1}|^2_{se}$ (green open downward triangles), $|v_{-1}|^2_{se}$ (violet dots).}
\label{figaferroevecii}
\end{centering}
\end{figure*}%

\paragraph{Collective excitations:}  In Fig.~\ref{figaferroeval}, we report the eigenvalue spectrum for the first point considering coupling parameters ($k_{L}, \Omega$) = ($2.0, 0.0$), with interaction strengths $c_{0} = 50$, $c_{2} = 2.5$. In Fig.~\ref{figaferroeval}(a), we present the excitation spectrum along the quasi-momentum direction $q_{x}$. The eigenspectrum reveals multi-band instabilities in the low-lying branch, with locations and amplitudes ($q_{x}, \omega$) = ($1.29, 1.578$), ($2.70, 0.920$), and ($4.50, 0.715$). Among the three imaginary frequencies, the third band represents roton instability, where $\text{Re}(\omega_{ll}) \approx 0$ and $\text{Im}(\omega_{ll}) \neq 0$~\cite{Wilson2012,Ravisankar2025}. In the first-excited branch of the eigenspectrum, we obtain two-band instability at $q_{x} =0.67, 2.70$ with amplitude $\omega =0.352, 0.920$. The second instability band emerges as a result of overlap between the low-lying and first-excited branch of the eigenspectrum, which is also discussed above in ferromagnetic interactions, and we define it as an unstable avoided crossing. The first instability band appears as a result of the overlap between the first-excited and second-excited branches of the eigenspectrum. Hence, the second-excited branch exhibits only a single instability band, which is located at $q_{x} = 0.67$, with amplitude $\omega = 0.352$. The overlap between the first-excited and second-excited branches is defined as a double unstable avoided crossing~\cite{Kronjager2010,Gangwar2025}. The presence of the instability bands in the eigenspectrum is symmetric about $q_{x}$.%

In Fig.~\ref{figaferroeval}(b), we present the collective excitation spectrum along the quasi-momentum direction $q_{y}$. We obtain only real eigenfrequencies; however, the gap among the branches closes at several points. The gap between the low-lying and first-excited branch closes at $q_{y} = -1.44$ and $3.16$, while it closes between the first-excited and second-excited at $q_{y} = -0.43$ and $1.26$. The low-lying branch develops a negative eigenfrequency with the magnitude of $\omega = -5.95$ at $q_{y} = -3.45$, which is responsible for the energetic instability of the condensate~\cite{Ozawa2013}.%

In Fig.~\ref{figaferroevecii}, we present the eigenvectors corresponding to the eigenvalue spectrum given in Fig.~\ref{figaferroeval}. In the top row, we exhibit the eigenvectors corresponding to Fig.~\ref{figaferroeval}(a) along the quasi-momentum direction $q_{x}$, and in the bottom row corresponding to Fig.~\ref{figaferroeval}(b) along the quasi-momentum direction $q_{y}$. Along $q_{x}$, in Fig.~\ref{figaferroevecii}(a i), the eigenvectors of the low-lying and first-excited branches exhibit the spin-like mode and satisfy the criterion given in Eq.~\ref{eqb:spin}, owing to the presence of the imaginary eigenfrequencies in the eigenspectrum. The low-lying branch exhibits the spin-like mode for the quasi-momentum range $q_{x} \in [0.0, 1.87], [2.37, 3.04]$, $[4.27, 4.72]$, and the first-excited branch depicts the spin-like mode $q_{x} \in [0.56, 0.77], [2.37, 3.04]$. At the point of unstable avoided crossing $q_{x} \in [2.37, 3.04]$, the eigenvector components are out of phase among themselves as well as with the other involved branch. In Fig.~\ref{figaferroevecii}(a ii), the zeroth component of the low-lying and first-excited branches exhibits the density-like modes independently. Since the eigenspectrum shows the presence of imaginary eigenfrequencies in the second-excited branch as a result of the unstable avoided crossing with the first-excited, we report the eigenvectors of both branches. In Fig.~\ref{figaferroevecii}(a iii), at the point of unstable avoided crossing for quasi-momentum range $q_{x} \in [0.56, 0.77]$, the eigenvector components exhibit the spin-like mode, due to the presence of the imaginary eigenfrequency in the eigenspectrum, which follows the criterion,   
\begin{align} \label{eqc:spin}
& \vert u_{+1} \vert_{se}^{2} - \vert u_{-1} \vert_{se}^{2} \neq 0, \quad
\vert v_{+1} \vert_{se}^{2} - \vert v_{-1} \vert_{se}^{2} \neq 0, \nonumber \\
& \vert u_{+1} \vert_{fe}^{2} - \vert u_{-1} \vert_{fe}^{2} \neq 0, \quad
\vert v_{+1} \vert_{fe}^{2} - \vert v_{-1} \vert_{fe}^{2} \neq 0.
\end{align}
In Fig.~\ref{figaferroevecii}(a iv), the zeroth component of the low-lying and first-excited branches exhibits the density-like modes independently. Along the $q_{y}$ direction, in Fig.~\ref{figaferroevecii}(b i), a flip occurs in the eigenvector components at the point of avoided crossings at $q_{y} = -1.46$ and $q_{y} = 3.16$ for both low-lying and first-excited branches of the eigenspectrum~\cite{Abad2013}. However, due to the presence of real eigenfrequencies, the eigenvector components exhibit the density-like mode, following the criterion given in Eq.~\ref{eqb:density}. In Fig.~\ref{figaferroevecii}(b ii), the zeroth component of the eigenvector depicts the density-like mode independently. Similar to the $q_{x}$ direction, here also, we report eigenvectors of the second-excited and first-excited branches separately, owing to the avoided crossing between them. A flip occurs in the eigenvector components at the point of avoided crossing $q_{y} = -0.43$, and $1.26$ [Fig.~\ref{figaferroevecii}(b iii)] for both first-excited and second-excited branches of the eigenspectrum~\cite{Abad2013}. However, due to the presence of real eigenfrequencies only, the eigenvector components exhibit the density-like mode, following the relation,
\begin{align} \label{eqc:density}
& \vert u_{+1} \vert_{se}^{2} - \vert u_{-1} \vert_{se}^{2} = 0, \quad
\vert v_{+1} \vert_{se}^{2} - \vert v_{-1} \vert_{se}^{2} = 0, \nonumber \\
& \vert u_{+1} \vert_{fe}^{2} - \vert u_{-1} \vert_{fe}^{2} = 0, \quad
\vert v_{+1} \vert_{fe}^{2} - \vert v_{-1} \vert_{fe}^{2} = 0.
\end{align}
The zeroth component of the eigenvector depicts the density-like mode independently [Fig.~\ref{figaferroevecii}(b iv)].%

\paragraph{Numerical Simulation:}

\begin{figure*}[!htb] 
\begin{centering}
\centering\includegraphics[width=0.99\linewidth]{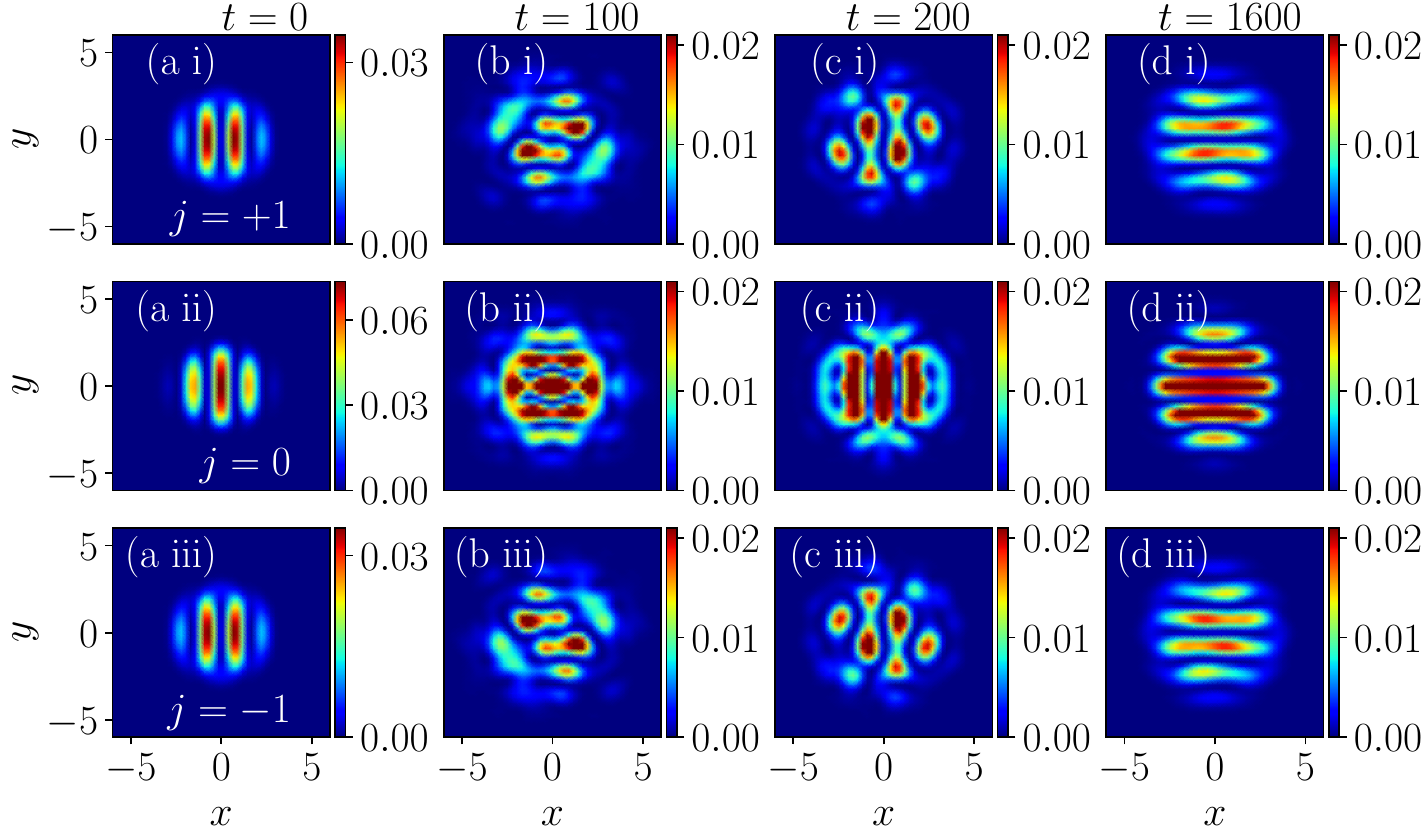}
\caption{Pseudocolor plots of the condensate density for the spin-1 system. Plot of (a) the ground state density profile exhibiting a superstripe phase and (b--d) density profiles during dynamical evolution at times $t = 100$, $200$, and $1600$ (in arbitrary units), respectively. In each subfigure, columns (i), (ii), and (iii) correspond to the $m_F = +1$, $0$, and $-1$ spinor components. The system parameters are fixed at $k_{L} = 2.0$, $\Omega = 0.0$, and interaction strengths $c_0 = 50$, $c_2 = 2.5$.}
\label{figaferronumi}
\end{centering}
\end{figure*}
 
\begin{figure}[!htb] 
\begin{centering}
\centering\includegraphics[width=0.99\linewidth]{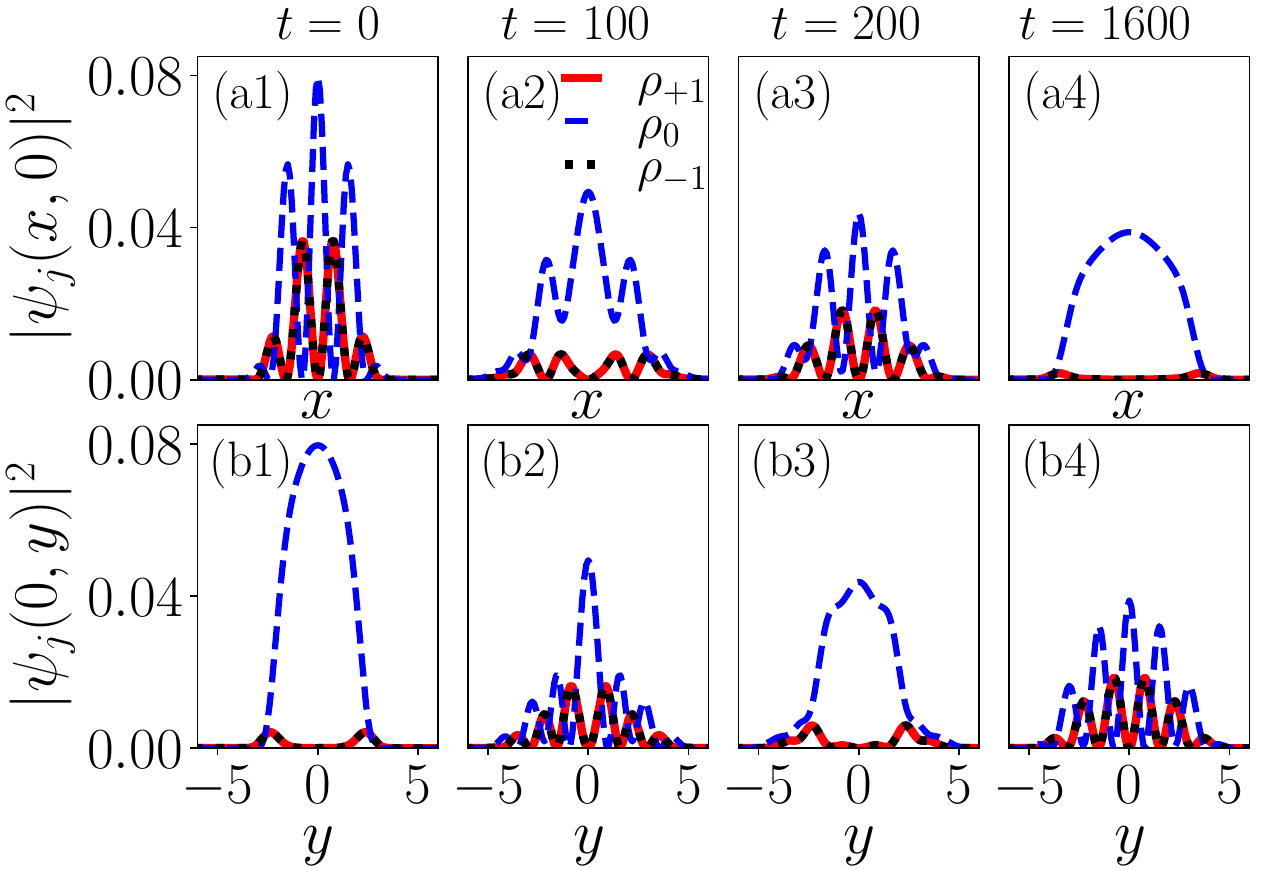}
\caption{(a1, b1) Ground-state density profiles. (a2--a4, b2--b4) Density profiles at evolution times $t = 100$, $200$, and $1600$ (arb.~units). The top row shows the density along the spatial $x$-direction at $y = 0$, and the bottom row shows the density along the $y$-direction at $x = 0$. Coupling parameters and interaction strengths are the same as in Fig.~\ref{figaferronumi}. The solid red line represents $\vert \psi_{+1} \vert^{2}$, the dashed blue line $\vert \psi_{0} \vert^{2}$, and the dotted black line $\vert \psi_{-1} \vert^{2}$.}
\label{figaferronumii}
\end{centering}
\end{figure}

In this section, we present the numerical simulation generating ground state density profile and performing quench dynamics of the condensate for coupling strengths ($k_{L}, \Omega$) = ($2.0, 0.0$), with interaction strengths $c_{0} = 50$, $c_{2} = 2.5$.%

In Fig.~\ref{figaferronumi}(a i-a iii), we show the ground state density profile, which corresponds to a superstripe phase in a Rashba-type SO coupled condensate. All three spinor components namely $m_{F} = +1, 0, -1$, participate in this phase. The $m_{F} = \pm 1$ components exhibit overlapping behaviour, while the $m_{F} = 0$ fill in the maxima and minima, leading to observed density modulation~\cite{Adhikari_2021, Adhikarimultr_2021}. To perform the quench dynamics, we reduced the trap strength ($\lambda$) to half ($\lambda/2$) of its original value. In Fig.~\ref{figaferronumi}(b i-b iii), at t = 100 units, small modulations appear inside the condensate density, and the profile starts to split into several bright spots. At $t = 200$ units, the modulated pattern becomes stronger. The originally smooth shape is now broken into multiple localized peaks [Fig.~\ref{figaferronumi}(c i-c iii)]. At a later time, $t = 1600$ units, in Fig.~\ref{figaferronumi}(d i-d iii), the state has fully reorganized, with an increased number of stripes in the zeroth component of the condensate. This indicates that the unstable modes have evolved, altering the overall configuration.%

In Fig.~\ref{figaferronumii}, we report the line plots of the density profiles shown in Fig.~\ref{figaferronumi}. The top row presents the profile along the spatial $x$-direction at $y=0.0$, while the bottom row depicts it along the $y$-direction at $x=0.0$. Along the $x$-direction, Fig.~\ref{figaferronumii}(a1) shows the ground-state density profile, which corresponds to a superstripe phase with density modulations produced by all three spinor components $m_{F} = +1, 0, -1$. During the time evolution, the density begins to distort; the superstripe pattern disappears, forming a stripe-like phase by $t = 100$ units [Fig.~\ref{figaferronumii}(a2)]. At $t = 200$ units, the condensate expands spatially, exhibiting an increased number of stripes in the $m_F=0$ component while maintaining stripes in all three components [Fig.~\ref{figaferronumii}(a3)]. At a later time, $t = 1600$ units, the $m_F = \pm 1$ components vanish, while the $m_F=0$ component develops a single-peak structure [Fig.~\ref{figaferronumii}(a4)]. Along the $y$-direction, Fig.~\ref{figaferronumii}(b1) shows the ground-state profile, where the $m_F=0$ component has a single-peak structure, and the $m_F = \pm 1$ components exhibit two symmetric humps. During the dynamics, a stripe-type phase emerges at $t = 100$ units [Fig.~\ref{figaferronumii}(b2)], which further evolves into a filament-like shape in the $m_F=0$ component, while the $m_F = \pm 1$ components retain symmetric humps at $t = 200$ units [Fig.~\ref{figaferronumii}(b3)]. By $t = 1600$ units, a stripe phase appears in all components, with an increased number of stripes in the $m_F=0$ component. These line plots confirm the alteration of the condensate configuration due to the growth of unstable modes.

We also analyze the total energy of the condensate in the antiferromagnetic regime for the interaction strengths $c_{0} = 50$, $c_{2} = 2.5$. We find that for $k_{L} = 2.0$ and $\Omega=0$, during the time evolution, the total energy initially has a value of $E = -0.26$ at $t = 0$ units. It begins to decrease following the abrupt change of the trap strength and reaches $E = -0.60$ at $t = 512.50$ units. Beyond this point, the energy of the condensate gets saturated (not shown here), implying the dynamical stability of the condensate.

Another noteworthy aspect arises in the antiferromagnetic regime discussed in the Appendix~\ref{afm:secp}, where we analyze the collective excitation in conjunction with a real-time scheme. Although the qualitative features of the excitation, such as the presence of double unstable avoided crossings and stable avoided crossings [see Figs.~\ref{figaferroevalii} and ~\ref{figaferroeveciib}], remain similar to those observed in the ferromagnetic case, the associated instability amplitude and the range of quasi-momenta over which they occur are significantly enhanced. As a consequence of these stronger instabilities, the condensate dynamics exhibit pronounced density fragmentation at early times, which subsequently evolves into the configuration characterized by the emergence of two dominant density lobes~[see Figs.~\ref{figaferronumiii} and ~\ref{figaferronumiv}].  


\section{Summary and Conclusions }
\label{sec:summ}

In this work, we have investigated the collective excitation spectrum and dynamical stability of quasi-two-dimensional Rashba spin-orbit and Rabi-coupled spin-1 Bose-Einstein condensates. Using Bogoliubov-de Gennes analysis complemented by quench dynamics, we systematically analyzed the interplay between spin-orbit coupling and interaction for the ferromagnetic and antiferromagnetic cases.

We first examined the single-particle spectrum and showed that the Rabi coupling opens gaps between the three branches of the excitation spectrum, with the gap magnitude set by $\Omega$ between $\omega_{\pm 1}$ and $\omega_{0}$ and $2\Omega$ between $\omega_{-}$ and $\omega_{+}$. The lowest energy ($\omega_{-}$) branch undergoes a transition from a single minimum to a double minimum structure when $\Omega < k_L^2$. While this transition generally occurs along the $q_x$ direction, along the $q_y$ direction it requires the simultaneous presence of both spin–orbit and Rabi couplings, highlighting their nontrivial interplay. 

For ferromagnetic interactions, the collective excitation spectrum reveals a rich structure in the $k_L-\Omega$ parameter space, which can be classified into three distinct regions characterized by their dynamical and energetic stability. Region I is fully stable and supports well-defined phonon modes with density-like character. Regions II and III exhibit various forms of dynamical instabilities, including single-band and multi-band instabilities, unstable avoided crossings, and maxon–roton features. These instabilities are accompanied by characteristic changes in the eigenvector structure, including density–spin mixing, spin-like modes, and mode flipping at avoided crossings. Notably, Region III differs from Region II by supporting instabilities that coexist with phonon modes in the low-lying and first-excited branches.

We also analyzed the collective excitations and dynamics for antiferromagnetic interactions. In this case, the spectrum exhibits multiple unstable avoided crossings involving low-lying, first-excited, and second-excited branches, leading to both multi-band and single-band instabilities. The corresponding eigenmodes are predominantly spin-like and display characteristic phase reversals at avoided crossings. Dynamically, in the absence of Rabi coupling, the system supports a supersolid-like ground state with density modulations arising from all three spin components. With increasing spin–orbit coupling, instabilities strongly modify the density profile, leading to enhanced stripe formation at intermediate coupling strengths and fragmentation into spatially separated lobes at larger coupling.

Taken together, the analysis performed in our work establishes a unified picture of how Rashba spin-orbit coupling, Rabi coupling, and spin-dependent interaction jointly shape the excitation spectrum and non-equilibrium dynamics of the spin-1 condensates. Across both the ferromagnetic and antiferromagnetic interactions, dynamical instabilities are shown to be intimately connected to the structure of avoided crossings in the excitation spectrum and to the evolving characteristics of the corresponding eigenmodes. The real-time dynamics consistently indicate these spectral features, with unstable modes seeding the pattern formation, density modulation, and fragmentation on experimentally accessible timescales. In particular, for the antiferromagnetic cases, we have demonstrated how the enhanced instability amplitude and extended unstable momentum ranges can lead to the emergence of a supersolid-like structure.

The present work suggests several promising directions for future research. First, exploring different physical geometries and interactions, such as three-dimensional confinement, anisotropic spin-orbit coupling~\cite{Li:2012}, or the inclusion of long-range dipolar interactions~\cite{Lewkowicz:2025}, may reveal new instability mechanisms and emergent orders, including supercurrent-carrying supersolid phases~\cite{PhysRevResearch.6.023048}. Second, employing Floquet engineering to make the spin-orbit and Rabi couplings time-dependent could provide a powerful route for dynamically controlling nonequilibrium phases and probing their experimental signatures in spin-1 Bose-Einstein condensates~\cite{Goldman:2014}. Finally, examining the influence of effects beyond the current zero-temperature mean-field model, particularly three-body interactions~\cite{Ravisankar2025} and finite-temperature fluctuations~\cite{roy2025thermal}, on the collective excitation spectrum and phase stability would be crucial for connecting these theoretical findings to realistic experimental conditions.

\acknowledgments 
We gratefully acknowledge our Param-Ishan and Param Kamrupa supercomputing facility (IITG), where all numerical simulations were performed. S.K.G gratefully acknowledges a research fellowship from MoE, Government of India. P.M. acknowledges support from the MoE, through the RUSA 2.0 (Bharathidasan University -- Physical Sciences).

\appendix
\counterwithin{figure}{section}

\section{Relevant terms of the BdG matrix of collective excitations}
\label{matrx:BdG}

In this appendix, we provide an explicit form of the matrix elements of the BdG matrix Eq.~(\ref{bdgmatrix}) used in Section~\ref{sec:collexc}. 
\onecolumngrid
The matrix elements of Eq. ~(\ref{bdgmatrix}) read as
\begin{subequations}
 \begin{align}
 H_{+} = & \frac{q_{x}^{2} + q_{y}^{2}}{2} + c_{0}(2\phi_{+1}^{2} + \phi_{0}^{2} +\phi_{-1}^{2}) + c_{2}(2\phi_{+1}^{2} + \phi_{0}^{2} -\phi_{-1}^{2}) \\
 H_{0} = & \frac{q_{x}^{2} + q_{y}^{2}}{2} + c_{0}(\phi_{+1}^{2} + 2\phi_{0}^{2} +\phi_{-1}^{2}) + c_{2}(\phi_{+1}^{2} +\phi_{-1}^{2})
\\
H_{-} = & \frac{q_{x}^{2} + q_{y}^{2}}{2} + c_{0}(\phi_{+1}^{2} + \phi_{0}^{2} +2\phi_{-1}^{2}) + c_{2}(2\phi_{-1}^{2} + \phi_{0}^{2} -\phi_{+1}^{2})
\\
\mu_{+}\phi_{+1} = & c_{0}(\phi_{+1}^{2} + \phi_{0}^{2} +\phi_{-1}^{2}) \phi_{+1} + c_{2}(\phi_{+1}^{2} + \phi_{0}^{2} -\phi_{-1}^{2}) \phi_{+1} + c_{2} \phi_{0}^{2} \phi_{-1}^{*} +  \frac{\Omega}{\sqrt{2}} \phi_{0}
\\
\mu_{0} \phi_{0}= & c_{0}(\phi_{+1}^{2} + \phi_{0}^{2} +\phi_{-1}^{2}) \phi_{0} + c_{2}(\phi_{+1}^{2} + \phi_{-1}^{2})\phi_{0} + 2 c_{2}\phi_{0}^{*}\phi_{+1}\phi_{-1} +  \frac{\Omega}{\sqrt{2}}(\phi_{+1}+\phi_{-1}) 
\\
\mu_{-} \phi_{-1} = & c_{0}(\phi_{+1}^{2} + \phi_{0}^{2} +\phi_{-1}^{2}) \phi_{-1} + c_{2}(\phi_{-1}^{2} + \phi_{0}^{2} -\phi_{+1}^{2}) \phi_{-1} + c_{2} \phi_{0}^{2} \phi_{+1}^{*} +  \frac{\Omega}{\sqrt{2}} \phi_{0}
\end{align}
\end{subequations} %
\begin{align}
  \mathcal{L}_{12} & = C^{+}\phi_{+1}^{2}; \;\; \mathcal{L}_{13}= C^{+}\phi_{0}^{*}\phi_{+1}+ \frac{k_{L}}{\sqrt{2}}(i q_{x} + q_{y})+ 2c_{2}\phi_{0}\phi_{-1}^{*}+ \frac{\Omega}{\sqrt{2}}; \;\; \mathcal{L}_{14}=C^{+}\phi_{0}\phi_{+1}; \;\; \mathcal{L}_{15}=C^{-}\phi_{-1}^{*}\phi_{+1} \notag \\
  \mathcal{L}_{16} & =C^{-}\phi_{-1}\phi_{+1} +c_{2} \phi_{0}^{2}; \;\; \mathcal{L}_{21}=-C^{+}\phi_{+1}^{*2}; \;\; \mathcal{L}_{23}=-C^{+}\phi_{0}^{*}\phi_{+1}^{*}; \;\; \mathcal{L}_{24} = -C^{+}\phi_{0}\phi_{+1}^{*} + \frac{k_{L}}{\sqrt{2}}(-i q_{x} + q_{y}) -2c_{2}\phi_{0}^{*}\phi_{-1}-\frac{\Omega}{\sqrt{2}}; \notag \\
  \mathcal{L}_{25} & =-C^{-}\phi_{-1}^{*}\phi_{+1}^{*} - c_{2}\phi_{0}^{*2}; 
  ;\;\ \mathcal{L}_{26}=-C^{-}\phi_{-1}\phi_{+1}^{*}; \;\; \mathcal{L}_{31} = C^{+}\phi_{+1}^{*}\phi_{0} + 2c_{2}\phi_{0}^{*}\phi_{-1} +\frac{k_{L}}{\sqrt{2}}(-i q_{x} + q_{y}) +\frac{\Omega}{\sqrt{2}}; \;\; \mathcal{L}_{32}=C^{+}\phi_{+1}\phi_{0}; \notag \\ \mathcal{L}_{34} & =c_{0}\phi_{0}^{2} +2 c_{2}\phi_{+1}\phi_{-1}; \;\;  \mathcal{L}_{35} = C^{+}\phi_{-1}^{*}\phi_{0} + 2c_{2}\phi_{0}^{*}\phi_{+1} +\frac{k_{L}}{\sqrt{2}}(i q_{x} + q_{y}) +\frac{\Omega}{\sqrt{2}}; \;\; \mathcal{L}_{36}=C^{+}\phi_{-1}\phi_{0}; \;\; \mathcal{L}_{41}=-C^{+}\phi_{+1}^{*}\phi_{0}^{*} \notag \\
  \mathcal{L}_{42} & = -C^{+}\phi_{+1}\phi_{0}^{*}+\frac{k_{L}}{\sqrt{2}}(i q_{x}+ q_{y})- 2c_{2}\phi_{0}\phi_{-1}^{*}-\frac{\Omega}{\sqrt{2}}; \;\; \mathcal{L}_{43}=-c_{0}\phi_{0}^{*2}-2 c_{2}\phi_{+1}^{*}\phi_{-1}^{*}; \;\; \mathcal{L}_{45}=-C^{+}\phi_{-1}^{*}\phi_{0}^{*}; \;\; \notag
  \mathcal{L}_{46} & = -C^{+}\phi_{-1}\phi_{0}^{*} + \frac{k_{L}}{\sqrt{2}} (-i q_{x}+q_{y})- 2c_{2}\phi_{0}\phi_{+1}^{*}- \frac{\Omega}{\sqrt{2}}; \;\; \mathcal{L}_{51}=C^{-}\phi_{+1}^{*}\phi_{-1}; \;\; \mathcal{L}_{52}=C^{-}\phi_{+1}\phi_{-1}+c_{2}\phi_{0}^{2}; \;\;  \notag \\
  \mathcal{L}_{53} & = C^{+}\phi_{0}^{*}\phi_{-1}+ \frac{k_{L}}{\sqrt{2}}(-i q_{x} + q_{y}) + 2c_{2}\phi_{0}\phi_{+1}^{*} + \frac{\Omega}{\sqrt{2}}; \;\; \mathcal{L}_{54}=C^{+}\phi_{0}\phi_{-1}; \;\; \mathcal{L}_{56}=C^{+}\phi_{-1}^{2}; \;\; \mathcal{L}_{61}= -C^{-}\phi_{+1}^{*}\phi_{-1}^{*}-c_{2}\phi_{0}^{*2}; \notag \\
  \mathcal{L}_{62} & =- C^{-}\phi_{+1}\phi_{-1}^{*}; \;\; \mathcal{L}_{63}=-C^{+}\phi_{0}^{*}\phi_{-1}^{*}; \;\; \mathcal{L}_{64} = -C^{+}\phi_{0}\phi_{-1}^{*}+\frac{k_{L}}{\sqrt{2}}(i q_{x}+ q_{y})-2c_{2}\phi_{0}^{*}\phi_{+1}-\frac{\Omega}{\sqrt{2}}; \;\; \mathcal{L}_{65}= -C^{+} \phi_{-1}^{*2}. \notag
\end{align}
Also,
\begin{align}
 C^{+}\equiv c_{0} + c_{2}, \;C^{-}\equiv c_{0} - c_{2}   \notag
\end{align}
The coefficients for the BdG characteristic equation~(\ref{bdgex}) are given as follows:
\begin{align}
b= & 0, \\
c= &\frac{1}{16}\bigg( -4 (q_{x}^{2} + q_{y}^{2}) (3 (q_{x}^{2} + q_{y}^{2}) + 8 k_{L}^{2}) - 48 (q_{x}^{2} + q_{y}^{2})\Omega -80 \Omega^{2} + 4 c_{0}^{2} + 4 c_{2} (q_{x}^{2} + q_{y}^{2} + 26 \Omega) \notag \\
&-61 c_{2}^{2} -8 c_{0}(q_{x}^{2} + q_{y}^{2} - 2 \Omega + c_{2}) \bigg) \\
d= & \frac{1}{2} k_{L} q_{y} \bigg(-8 \Omega (q_{x}^{2} + q_{y}^{2} + 2 \Omega)-2 c_{0}^{2} +c_{0}(4(q_{x}^{2} + q_{y}^{2} + \Omega) -3 c_{2}) +(12 (q_{x}^{2} + q_{y}^{2})+ 26 \Omega -7 c_{2})c_{2} \bigg) 
\end{align}
\begin{align}
e= &\frac{1}{32}\bigg( 2 (3 (q_{x}^{2} + q_{y}^{2})^{4} + 16 (q_{x}^{2} + q_{y}^{2})^{2} k_{L}^{4} + 24 (q_{x}^{2} + q_{y}^{2})^{3} \Omega + 8 (9 (q_{x}^{2} + q_{y}^{2})^{2} - 4 (q_{y} - q_{x}) (q_{y} + q_{x}) k_{L}^{2} ) \Omega^{2} \notag \\
&+ 96 (q_{x}^{2} + q_{y}^{2}) \Omega^{3} +64 \Omega^{4} )
+ 4c_{0}^{3}(q_{x}^{2} + q_{y}^{2} + 2 \Omega - 2c_{2}) -4 ((q_{x}^{2} + q_{y}^{2})^{3} - 2 (q_{y}^{2} -5 q_{x}^{2}) (q_{x}^{2} + q_{y}^{2}) k_{L}^{2} \notag \\ 
&+ 6 (q_{x}^{2} + q_{y}^{2})^{2} \Omega + 4(-13 q_{y}^{2} +17 q_{x}^{2}) k_{L}^{2} \Omega + 10 (q_{x}^{2} + q_{y}^{2}) \Omega^{2} + 52 \Omega^{2})c_{2}
- (15 q_{y}^{4} + 15 q_{x}^{4} -236 \Omega^{2} + 4 q_{x}^{2}(-32 k_{L}^{2} +29 \Omega) \notag \\
&+ 2q_{y}^{2}(15 q_{x}^{2} + 52 k_{L}^{2} + 58 \Omega)) c_{2}^{2} + (71 (q_{x}^{2} + q_{y}^{2}) - 210 \Omega ) c_{2}^{3} + 96 c_{2}^{4}
- 4 c_{0}^{2}( 3 q_{y}^{4} + (q_{x}^{2} + 4 \Omega) (3 q_{x}^{2} + 4 \Omega)\notag \\
&+ 2 q_{y}^{2} (3 q_{x}^{2} - 2k_{L}^{2} + 8 \Omega) + 2c_{2}(-6(q_{x}^{2} + q_{y}^{2} + 2\Omega) + 5 c_{2})) + c_{0}(8((q_{x}^{2} + q_{y}^{2})^{3} - 2 (q_{y}^{2} - 2 q_{x}^{2}) (q_{x}^{2} + q_{y}^{2})k_{L}^{2} \notag \\
&+2 (3(q_{x}^{2} + q_{y}^{2})^{2} + 2 q_{y}^{2} k_{L}^{2})\Omega + 10(q_{x}^{2} + q_{y}^{2}) \Omega^{2} - 4 \Omega^{3}) + c_{2} (-4 (9 q_{y}^{4} + 9q_{x}^{4} + 42 q_{x}^{2} \Omega - 6 \Omega^{2} + 2 q_{y}^{2} (9 q_{x}^{2} + 5 k_{L}^{2} + 21 \Omega)) \notag \\
&+ c_{2}(85 (q_{x}^{2} + q_{y}^{2}) -22 \Omega +14 c_{2}))) \bigg)\\
f= &\frac{1}{8} q_{y} k_{L}\bigg( 8 \Omega (q_{x}^{2} + q_{y}^{2} + 2\Omega)^{3} + 2 (q_{x}^{2}+ q_{y}^{2} + 2 \Omega)^{2} c_{0}^{2} -2(6 (q_{x}^{2} + q_{y}^{2})^{3} + (37 (q_{x}^{2} + q_{y}^{2})^{2} + 8 q_{x}^{2}k_{L}^{2})\Omega + 76(q_{x}^{2} + q_{y}^{2}) \Omega^{2} \notag \\
&+ 52\Omega^{3})c_{2} + (7q_{y}^{4} + 7 q_{x}^{4} + 28 \Omega^{2} + 14 q_{y}^{2}(q_{x}^{2} + 2 \Omega) + 4q_{x}^{2}(8k_{L}^{2} + 7 \Omega))c_{2}^{2} - c_{0} (4 ((q_{x}^{2} + q_{y}^{2})^{3} + (5(q_{x}^{2} + q_{y}^{2})^{2} - 4q_{x}^{2} k_{L}^{2})\Omega \notag \\ 
&+ 8 (q_{x}^{2} + q_{y}^{2})\Omega^{2} + 4 \Omega^{2}) - 3 (q_{y}^{2} + q_{x}^{2} + 2\Omega)^{2} c_{2}) \bigg)\\
g= & \frac{1}{256} \bigg( -8 (q_{x}^{2} + q_{y}^{2} + 2 \Omega)^{2} c_{0}^{3} (q_{x}^{2} + q_{y}^{2} + 2 \Omega - 2c_{2}) + 4 (q_{x}^{2} + q_{y}^{2} + 2 \Omega)c_{0}^{2} ((q_{x}^{2} + q_{y}^{2} + 2\Omega) (5q_{x}^{4} + 5q_{y}^{4}+ 28 \Omega^{2} \notag \\
&+ 4 q_{x}^{2}(-4k_{L}^{2} + 7\Omega) + 2q_{y}^{2}(5 q_{x}^{2} - 4 k_{L}^{2} + 14 \Omega)) - 8(3 q_{y}^{4} + 3 q_{x}^{4} - 4 q_{x}^{2}(k_{L}^{2} - 3\Omega) + 12 \Omega^{2} + 6q_{y}^{2}(q_{x}^{2} + 2\Omega))c_{2} \notag \\
&+ 20(q_{x}^{2} + q_{y}^{2} + 2\Omega)c_{2}^{2}) - 2c_{0}(4(q_{x}^{2} + q_{y}^{2}+ 2\Omega)((q_{x}^{2} + q_{y}^{2})(q_{y}^{2} + q_{x}(q_{x} - 2 k_{L}))(q_{x}^{2} + q_{y}^{2} - 4k_{L}^{2}) (q_{y}^{2} + q_{x} (q_{x} + 2 k_{L})) \notag \\ 
&+ 12(q_{x}^{2} + q_{y}^{2}) (q_{y}^{2} + q_{x} (q_{x} - 2k_{L}))(q_{y}^{2} + q_{x}(q_{x} + 2 k_{L}))\Omega + 4(11 (q_{x}^{2} + q_{y}^{2})^{2} + 4 (q_{y}^{2} - 3q_{x}^{2})k_{L}^{2})\Omega^{2} + 48 (q_{x}^{2} + q_{y}^{2})\Omega^{3}) \notag \\
&+ c_{2}(8(-5 (q_{x}^{2} + q_{y}^{2})^{4} - (5q_{y}^{2}- 22 q_{x}^{2})(q_{x}^{2} + q_{y}^{2})^{2}k_{L}^{2} - 16 q_{x}^{4}k_{L}^{4}- (q_{x}^{2} + q_{y}^{2}) (43(q_{x}^{2} + q_{y}^{2})^{2}+ 4(5q_{y}^{2} - 22 q_{x}^{2})k_{L}^{2})\Omega \notag \\
&- (111 (q_{x}^{2} + q_{y}^{2})^{2} + 4(5 q_{y}^{2}- 22 q_{x}^{2})k_{L}^{2})\Omega^{2}- 88 (q_{x}^{2} + q_{y}^{2})\Omega^{3} + 4\Omega^{4}) + (q_{x}^{2} + q_{y}^{2} + 2\Omega)c_{2} (85 (q_{x}^{2} + q_{y}^{2})^{2} - 160 q_{x}^{2} k_{L}^{2}\notag \\
&+ 148 (q_{x}^{2} + q_{y}^{2})\Omega -44 \Omega^{2} + 14(q_{x}^{2} + q_{y}^{2} + 2\Omega)c_{2}))) + (q_{x}^{2} + q_{y}^{2} + 2\Omega) (-4 (q_{x}^{2} + q_{y}^{2} + 2\Omega)((q_{x}^{2} + q_{y}^{2})(q_{x}^{2} + q_{y}^{2} - 4k_{L}^{2}) \notag \\
&+ 4 (q_{x}^{2} + q_{y}^{2} - 2q_{y} k_{L})\Omega) ((q_{x}^{2} + q_{y}^{2})(q_{x}^{2} + q_{y}^{2} - 4k_{L}^{2}) + 4 (q_{x}^{2} + q_{y}^{2} - 2q_{y} k_{L})\Omega) \notag \\
&+ c_{2}(4(q_{x}^{2} + q_{y}^{2})(q_{x}^{2} + q_{y}^{2} - 4k_{L}^{2})((q_{x}^{2} + q_{y}^{2})^{2}-8q_{x}^{2}k_{L}^{2})- 64 (q_{x}^{2} + q_{y}^{2}) ((q_{x}^{2} + q_{y}^{2})^{2}+ (7q_{y}^{2} - 3q_{x}^{2})k_{L}^{2})\Omega \notag \\ 
&- 16 (31 (q_{x}^{2} + q_{y}^{2})^{2} + 4 (13 q_{y}^{2} -11 q_{x}^{2})k_{L}^{2})\Omega^{2} -704 (q_{x}^{2} + q_{y}^{2})\Omega^{3} + c_{2}(91 (q_{x}^{2} + q_{y}^{2})^{3} +16 (13 q_{y}^{2} - 22 q_{x}^{2}) (q_{x}^{2} + q_{y}^{2})k_{L}^{2} \notag \\
&+ 658(q_{x}^{2} + q_{y}^{2})^{2} \Omega + 32 (13 q_{y}^{2} - 38 q_{x}^{2}) k_{L}^{2}\Omega + 724 (q_{x}^{2} + q_{y}^{2})\Omega^{2} - 456 \Omega^{3} + 2c_{2}(-71 q_{y}^{4} - 142 q_{y}^{2} q_{x}^{2} \notag \\ 
&-71 q_{x}^{4} + 256 q_{x}^{2}k_{L}^{2} + 68(q_{x}^{2} + q_{y}^{2})\Omega + 420 \Omega^{2}- 96 (q_{x}^{2} + q_{y}^{2} + 2\Omega)c_{2}))))\bigg)
\end{align}

\twocolumngrid
\section{Collective Excitations for Antiferromagnetic interactions}
\label{afm:secp}

\begin{figure}[!htb] 
\begin{centering}
\centering\includegraphics[width=0.95\linewidth]{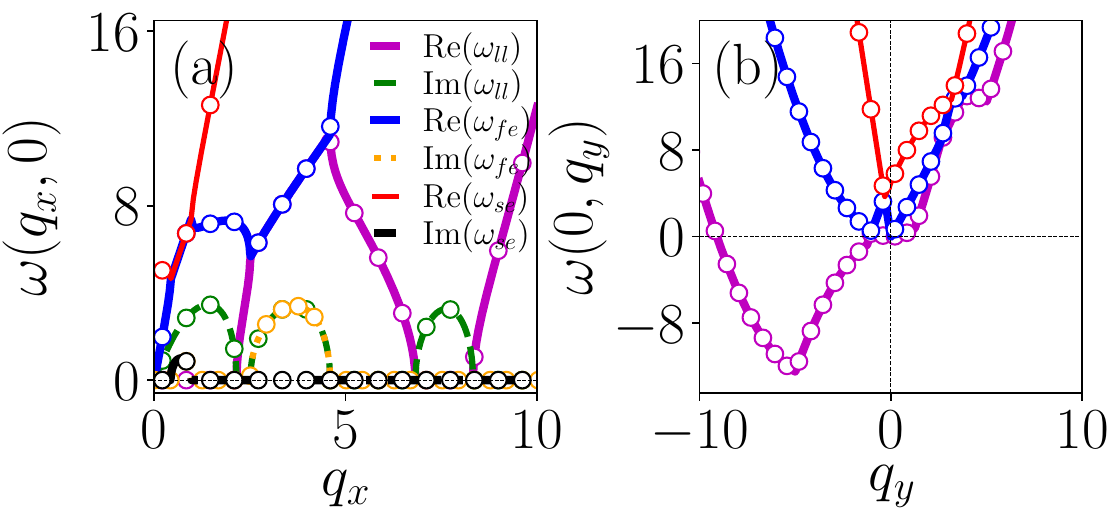}
\caption{Eigenvalue spectrum along the quasi-momentum directions (a) $q_x$ and (b) $q_y$. The coupling parameters are $k_L = 4.0$ and $\Omega = 0.0$, with interaction strengths $c_0 = 50$ and $c_2 = 2.5$. Line styles and symbols have the same meaning as in Fig.~\ref{figaferroeval}.}
\label{figaferroevalii}
\end{centering}
\end{figure}%

\begin{figure*}[!htb] 
\begin{centering}
\centering\includegraphics[width=0.99\linewidth]{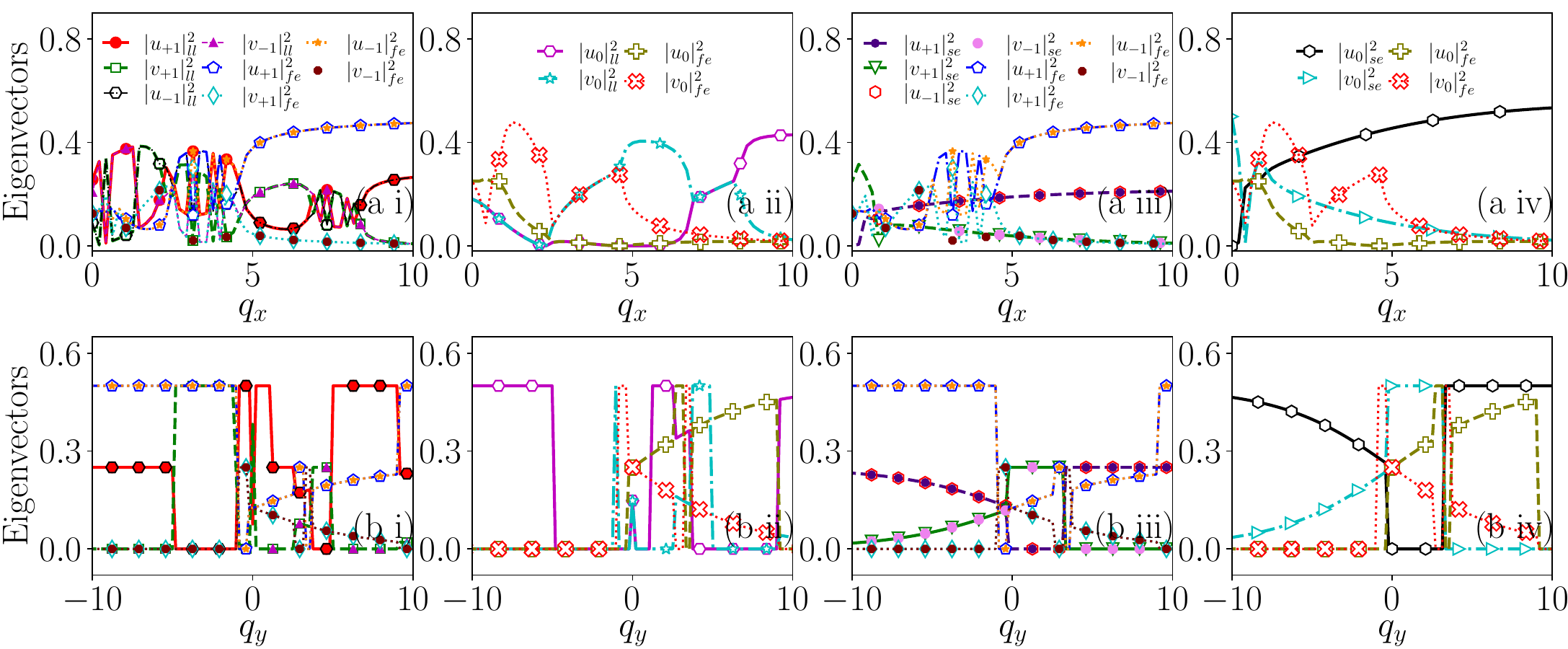}
\caption{Eigenvectors corresponding to Fig.~\ref{figaferroevalii}; (a i-a iv) along the quasi-momentum direction $q_{x}$, and (b i-b iv) along the quasi-momentum direction $q_{y}$. The lines and symbols have the similar meaning as in Fig.~\ref{figaferroevecii}.} 
\label{figaferroeveciib}
\end{centering}
\end{figure*}%

In this appendix, considering interaction strengths $c_{0} = 50$, $c_{2} = 2.5$ with SO coupling strength $k_{L} = 4.0$ in the absence of Rabi coupling ($\Omega \sim 0.0$), we present the collective excitations and corresponding ground state dynamics for the antiferromagnetic regime.  To scan the condensate behaviour for the above-mentioned point, we present the collective excitation spectrum and quench dynamics.%

\paragraph{Collective excitations:} Fig.~\ref{figaferroevalii} presents the collective excitation spectrum for SO coupling strength $k_{L} = 4.0$, in absence of Rabi coupling with interaction strengths $c_{0} = 50$, $c_{2} = 2.5$ along the quasi-momentum direction $q_{x}$, and $q_{y}$, separately. Along the quasi-momentum direction $q_{x}$, in Fig.~\ref{figaferroevalii}(a), we obtain three instability bands in the low-lying branch with location and amplitude ($q_{x}, \omega$) = ($1.42, 3.44$), ($3.62, 3.44$), and ($7.61, 3.27$). The second instability band arises from the overlap with the first-excited branch, exhibiting an unstable avoided crossing. At this point, the first-excited branch also develops an instability band of the same amplitude. In the first-excited branch, the second instability in the first-excited branch arises when it overlaps with the second-excited branch, located at $q_{x} = 0.68$, with amplitude $\omega = 1.01$. This band is shared by both the first- and second-excited branches and corresponds to the double unstable avoided crossing~\cite{Kronjager2010}. This is the only instability band that appears in the second excited branch of the eigenspectrum.    

Along quasi-momentum direction $q_{y}$, in Fig.~\ref{figaferroevalii}(b), we obtain only real eigenfrequencies in all three branches of the eigenspectrum. However, avoided crossings emerge among low-lying, first-excited, and second-excited branches of the eigenspectrum.  The avoided crossings between the low-lying and first-excited branches at $q_{y} = -1.02, 2.63$, $3.77$; between first-excited and second-excited branches at $q_{y} = -0.33, 3.14$. The low-lying branch develops the negative contribution to the excitation spectrum with amplitude $\omega = -12.54$ at $q_{y} = -5.05$. This negative excitation leads to energetic instability of the condensate~\cite{Ozawa2013}.%

In Fig.~\ref{figaferroeveciib}, we present the eigenvectors corresponding to the eigenvalue spectrum given in Fig.~\ref{figaferroevalii}. In the top row, we exhibit the eigenvectors along the quasi-momentum direction $q_{x}$, while in the bottom row, along the quasi-momentum direction $q_{y}$. Along $q_{x}$ direction, in Fig.~\ref{figaferroeveciib}(a i), we obtain spin-like modes in the low-lying and first-excited branches in eigenvector components following the criterion given in Eq.~\ref{eqb:spin}. In the low-lying branch, the eigenvectors depict the spinlike mode for quasi-momentum range $q_{x} \in [0, 2.14], [2.5, 4.60], [6.80, 8.33]$. For the first-excited branch, instability emerges at the point of overlap between the low-lying and first-excited branch, which is $q_{x} \in [2.5, 4.60]$ (second band in low-lying branch). At the point of unstable avoided crossing, the eigenvectors are out of phase with each other as well as with the other branch. The zeroth component of the eigenvectors exhibits the density-like mode independently [Fig.~\ref{figaferroeveciib}(a ii)]. Since the first and second excited branches show unstable avoided crossings in the eigenspectrum, we also present the corresponding eigenvectors for these two branches. In Fig.~\ref{figaferroeveciib}(a iii), the eigenvector components depict spin-like modes for quasi-momentum range $q_{x} \in [0.41, 0.98]$ in the first-excited and second-excited branches of the spectrum, following the criterion given in Eq.~\ref{eqc:spin}. At the point of unstable avoided crossings, the eigenvector components are out of phase within the branch as well as with other branches. The zeroth component of the eigenvectors depicts density-like mode independently [Fig.~\ref{figaferroeveciib}(a iv)]~\cite{Gangwar2025}.

Along the $q_{y}$ direction, in Fig.~\ref{figaferroeveciib}(b i), the eigenvector components of the low-lying and first-excited branches exhibit in-phase behavior (density-like modes), following the criterion given in Eq.~\ref{eqb:density}. A flip occurs in the eigenvector components at the point of avoided crossings between the branches, located at $q_{y} = -1.02, 2.63, 3.77$~\cite{Abad2013}. The zeroth component of the eigenvector depicts density-like mode independently [Fig.~\ref{figaferroeveciib}(b ii)]. Here, we also obtain the avoided crossings between the first-excited and second-excited branches of the eigenspectrum and report the eigenvectors separately. In Fig.~\ref{figaferroeveciib}(b iii), the eigenvector components of the first-excited and second-excited branches exhibit the density-like modes, following the criterion given in Eq.~\ref{eqc:density}. A flip appears in the eigenvector components at the point of avoided crossings between the branches at points $q_{y} = -0.33, 3.14$~\cite{Abad2013}. The zeroth component of the eigenvector depicts density-like mode independently [Fig.~\ref{figaferroeveciib}(b iv)].%

\begin{figure*}[!htb] 
\begin{centering}
\centering\includegraphics[width=0.99\linewidth]{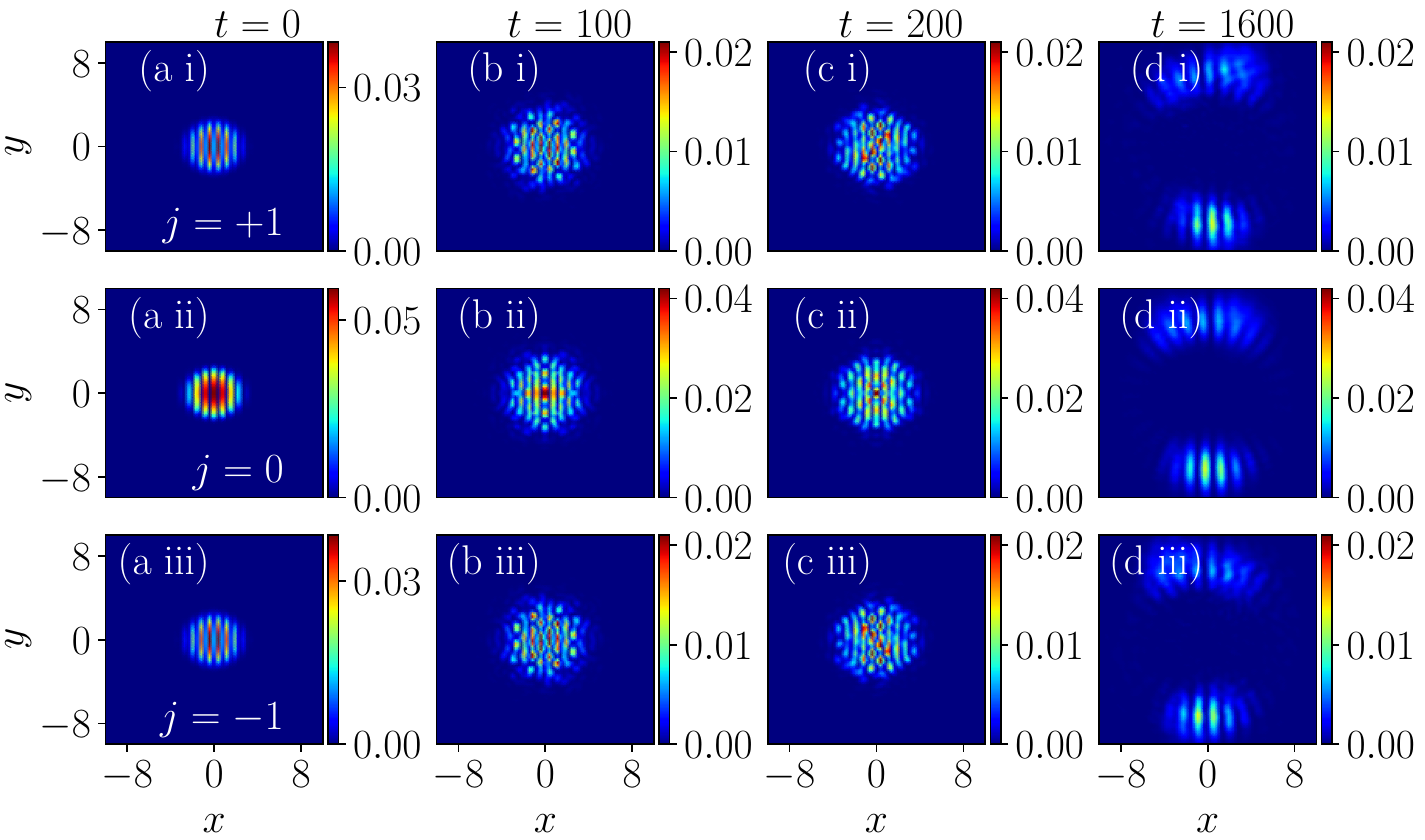}
\caption{Density profiles of a spin-1 condensate. (a) The initial ground-state profile, showing the superstripe phase. (b--d) Dynamical density profiles at evolution times $t = 100$, $200$, and $1600$ (in simulation units), respectively. For each state (a--d), the three columns (i), (ii), and (iii) correspond to the spinor components $m_F = +1$, $0$, and $-1$. The system parameters are $k_L = 4.0$, $\Omega = 0.0$, $c_0 = 50$, and $c_2 = 2.5$.}
\label{figaferronumiii}
\end{centering}
\end{figure*}%

\paragraph{Numerical Simulation:}
In this part of the appendix, we perform numerical simulation considering the SO coupling strength $k_{L} = 4.0$ in the absence of Rabi coupling ($\Omega = 0$) with interaction strengths $c_{0} = 50$, $c_{2} = 2.5$. Initially, we generate the ground state density profile, which is the superstripe wave phase [Fig.~\ref{figaferronumiii}(a i-a iii)]. Similar to the previous point regarding antiferromagnetic interactions, this superstripe wave phase also exhibits density modulation combining all three spinor components of the condensate. The $m_{F} = \pm 1$ components largely overlap, while the $m_{F} = 0$ component fills in the maxima and minima~\cite{Adhikari_2021,Adhikarimultr_2021}. To study the dynamics of the condensate, we quench the trap strength to half of its original value, i.e., $\lambda \rightarrow \lambda/2$. At t =100 units, small fluctuations start to appear on top of the main density peaks, marking the onset of fragmentation [Fig.~\ref{figaferronumiii}(b i-b iii)]. At $t = 200$ units, the density profile loses its clean shape. Fragmentation grows due to the growth of unstable modes in the condensate [Fig.~\ref{figaferronumiii}(c i-c iii)]. At a later time, at $t = 1600$ units, in Fig.~\ref{figaferronumiii}(b i-b iii), the condensate is fully fragmented, leaving only two vertical lobes in the density profile, which reflects the dynamically unstable nature of the condensate~\cite{Kronjager2010}.%

\begin{figure}[!htb] 
\begin{centering}
\centering\includegraphics[width=0.99\linewidth]{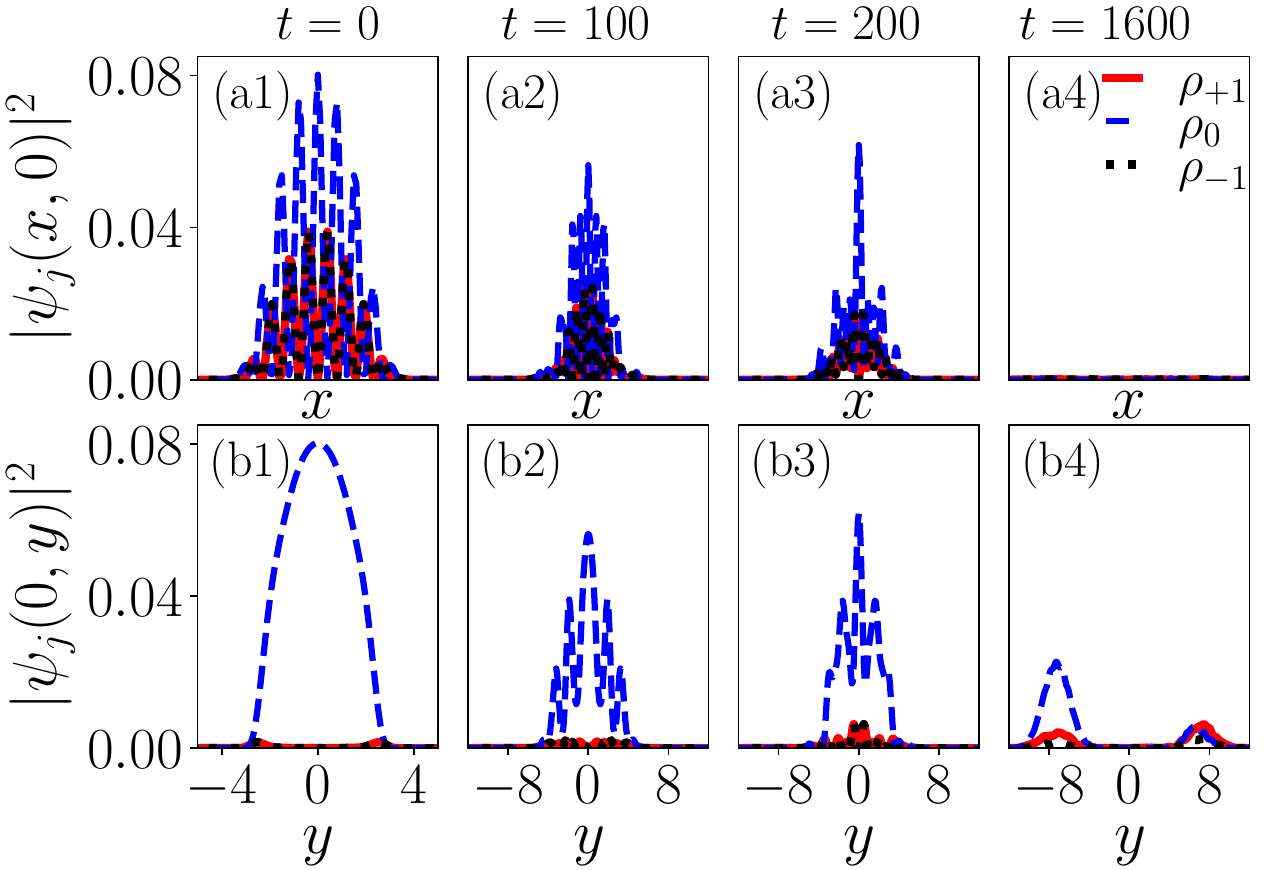}
\caption{(a) shows the ground-state density profile, while (b–d) display the density profiles during time evolution at $t = 100, 200,$ and $1600$ units, respectively. The top row shows densities along the $x$-direction at $y=0$, and the bottom row along the $y$-direction at $x=0$. The coupling parameters and interaction strengths are the same as in Fig.~\ref{figaferronumiii}. Line styles indicate the spinor components: solid red for $\vert \psi_{+1} \vert^{2}$, dashed blue for $\vert \psi_0 \vert^{2}$, and dotted black for $\vert \psi_{-1} \vert^{2}$.} %
\label{figaferronumiv}
\end{centering}
\end{figure}%

In Fig.~\ref{figaferronumiv}, we depict the line plots of the density profile corresponding to Fig.~\ref{figaferronumiii}. In the top row, we consider the variation along the spatial $x$-direction keeping $y = 0$, while in the bottom row, we present it along the spatial $y$-direction keeping $x = 0$. In the top row, in Fig.~\ref{figaferronumiv}(a1), we obtain the ground state density profile with the multi-peak density profile, where $m_{F} = \pm 1$ components overlap largely and zeroth components fill in for maxima and minima, producing a density modulation, which corresponds to the superstripe phase. During time evolution at $t = 100$ units, the density profile starts to deform and shrink in spatially [Fig.~\ref{figaferronumiv}(a2)]. Further, it keeps shrinking and fragmented at t= 200 units [Fig.~\ref{figaferronumiv}(a3)]. At a later time $t = 1600$ units in Fig.~\ref{figaferronumiv}(a4), all components of the density profile disappear along the spatial direction $x$. In the bottom row, in Fig.~\ref{figaferronumiv}(b1), we obtain the ground state along the spatial $y$-direction, where the $m_{F} = 0$ has a single peak density profile, while the $m_{F} = \pm 1$ depict the negligible two hump density profile. During the dynamics at $t = 100$ units, deformation begins, and all three components develop the multi-peak density profile [Fig.~\ref{figaferronumiv}(b2)]. Density profile keeps deforming, and the number of peaks reduces at $t = 200$ units [Fig.~\ref{figaferronumiv}(b3)]. At a later time $t = 1600$ units, in Fig.~\ref{figaferronumiv}(b4), all three components develop two hump density profile peaks at $y \approx 8$. The evolving behavior of the condensate density profile along spatial directions confirms the dynamically unstable nature of the condensate~\cite{Sadler2006, Kronjager2010}.%

The total energy of the condensate with anti-ferromagnetic interactions ($c_0=50$, $c_2=2.5$) for $k_{L} = 4.0$, in the absence of Rabi coupling, exhibits a gradual evolution following a quench in the trap strength. Starting from  $E = -6.26$ (in non-dimensional units) at $t = 0$, the energy relaxes to a lower value of $E=-6.5$ by $t = 213.50$ (data not shown), reflecting the adjustment of the condensate during its dynamical evolution.

\bibliography{reference.bib}

\end{document}